\documentclass[12pt]{article}
\usepackage{amsmath}
\usepackage{amssymb}
\usepackage{epsf}

\def\Tr{{\mathop{\mathrm{Tr}}\nolimits}}

\def\ev#1{{\langle{#1}\rangle}} 
\def\Ev#1{{\left\langle{#1}\right\rangle}} 

\def\CN{{\cal N}}
\def\CO{{\cal O}}

\def\Nh{{\widehat{N}}}

\def\Sh{{\widehat{S}}}

\def\zh{{\widehat{z}}}

\def\ft{{\widetilde{f}}}

\def\qt{{\widetilde{q}}}

\def\zt{{\widetilde{z}}}

\def\Nt{{\widetilde{N}}}

\def\Qt{{\widetilde{Q}}}

\def\unit{{1\kern-.65ex {\rm l}}}
\def\1{{1\kern-.65ex {\rm l}}}
\def\half{{\frac{1}{2}}}
\def\thalf{{\textstyle\frac{1}{2}}}

 

\makeatletter
\def\eqnarray{%
  \stepcounter{equation}\def\@currentlabel{\p@equation\theequation}%
  \global\@eqnswtrue \m@th \global\@eqcnt\z@ \tabskip\@centering
  \let\\\@eqncr
  $$\everycr{}\halign to\displaywidth\bgroup
    \hskip\@centering$\displaystyle\tabskip\z@skip{##}$\@eqnsel
   &\global\@eqcnt\@ne \hfil$\;{##}\;$\hfil
   &\global\@eqcnt\tw@ $\displaystyle{##}$\hfil\tabskip\@centering
   &\global\@eqcnt\thr@@ \hb@xt@\z@\bgroup\hss##\egroup
    \tabskip\z@skip
    \cr}
\makeatother

\def\La{\Lambda}
\def\nonu{\nonumber}
\newcommand{\bea}{\begin{eqnarray}}
\newcommand{\eea}{\end{eqnarray}}
 
\def\W #1{\widetilde{#1}}
\def\braket#1{\left\langle #1 \right\rangle}
\newcommand{\bean}{\begin{eqnarray*}}
\newcommand{\eean}{\end{eqnarray*}}
\def\a{{\alpha}}
\def\WH #1{\widehat{#1}}

\begin{document}
\begin{flushright}
TIT-HEP-519\\
UCLA/04/TEP/17\\
{\tt hep-th/0405101}\\
\end{flushright}

\vspace*{1.3cm} 
\centerline{\Large \bf Supersymmetric Gauge Theories with Flavors}
\vspace*{.3cm} 
\centerline{\Large \bf and Matrix Models }
\vspace*{1.5cm}
\centerline{{\bf Changhyun Ahn}$^{1,2}$, {\bf Bo Feng}$^1$,  
{\bf Yutaka Ookouchi}$^3$ and 
{\bf Masaki Shigemori}$^4$}
\vspace*{1.0cm} 
 \centerline{\it $^1$ School of Natural Sciences,
Institute for Advanced Study,
Olden Lane, Princeton NJ 08540, USA}
\centerline{\it $^2$Department of Physics,
Kyungpook National University, Taegu 702-701, Korea}
\centerline{\it $^3$Department of Physics,
 Tokyo Institute of
Technology, Tokyo 152-8511, Japan}
 \centerline{\it $^4$Department of Physics and Astronomy,
UCLA, Los Angeles, CA 90095-1547, USA}
\vspace*{0.8cm} 
\centerline{\tt
ahn@ias.edu,  \qquad fengb@ias.edu} 
\centerline{ \tt
ookouchi@th.phys.titech.ac.jp \qquad shige@physics.ucla.edu}
\vskip2cm

\centerline{\bf Abstract}
\vspace*{0.5cm}

\baselineskip=18pt

We present two results concerning the relation between poles and cuts by
using the example of $\CN=1$ $U(N_c)$ gauge theories with matter fields
in the adjoint, fundamental and anti-fundamental representations.
The first result is the on-shell possibility of poles, which are
associated with flavors and on the second sheet of the Riemann surface,
passing through the branch cut and getting to the first sheet.
The second result is the generalization of hep-th/0311181 (Intriligator,
Kraus, Ryzhov, Shigemori, and Vafa) to include flavors.  We clarify when
there are closed cuts and how to reproduce the results of the strong
coupling analysis by matrix model, by setting the glueball field to zero
from the beginning.
%
%
We also make remarks on the possible stringy explanations of the results
and on generalization to $SO(N_c)$ and $USp(2N_c)$ gauge groups.

%

\baselineskip=18pt
\newpage
\renewcommand{\theequation}{\arabic{section}\mbox{.}\arabic{equation}}


\setcounter{page}{1}

\section{Introduction}
\setcounter{equation}{0}
                                                                                                           
\baselineskip=18pt 

\indent
                                                                                                           
String theory can be a powerful tool to understand four dimensional
supersymmetric gauge theory which exhibits rich dynamics and allows an
exact analysis.  In \cite{CSW-II}, using the generalized Konishi anomaly
and matrix model \cite{DV}, ${\cal N}=1$ supersymmetric $U(N_c)$ gauge
theory with matter fields in the adjoint, fundamental and
anti-fundamental representations was studied.  The resolvents in the
quantum theory live on the two-sheeted Riemann surface defined by the
matrix model curve.
Their quantum behavior is characterized by the structure around the
branch cuts and poles, which are related to the RR flux contributions in
the Calabi--Yau geometry and flavor fields, respectively.  A pole
associated with flavor on the first sheet is related to the Higgs vacua
(corresponding to classical nonzero vacuum expectation value of the
fundamental) while a pole on the second sheet is related to the
pseudo-confining vacua where the classically vanishing vacuum
expectation value of the fundamental gets nonzero values due to quantum
correction.

It is known \cite{CSW-II} that Higgs vacua and pseudo-confining vacua,
which are distinct in the classical theory, are smoothly transformed
into each other in the quantum theory. This transition is realized on
the Riemann surface by moving poles located on the second sheet to pass
the branch cuts and enter the first sheet. This process was analyzed in
\cite{CSW-II} at the off-shell level by fixing the value of glueball fields during
the whole process.  However, in an on-shell process, the position of
poles and the width and position of branch cuts are correlated (when the
flavor poles are moved, the glueball field is also changed).  It was
conjectured in \cite{CSW-II} that for a given branch cut,
there is an upper bound for the number of poles (the number of flavors)
which can pass through the cut from the second sheet to the first sheet.

Our first aim of this paper is to confirm this conjecture and give the
corresponding upper bound for various gauge groups (in particular, we
will concentrate on the $U(N_c)$ gauge group). The main result is that
if $N_f \geq N_c $, the poles will not be able to pass through the cut
to the first sheet where $N_c$ is the effective fluxes associated with
the cut (and can be generalized to other gauge groups).

Another important development was made in \cite{IKRSV}, which was
inspired by \cite{Kraus:2003jf}.  In \cite{IKRSV}, which we will refer
to as IKRSV, it was shown that, to correctly compute the prediction of
string theory (matrix model), it is crucial to determine whether the
glueball is really a good variable or not.  A prescription was given,
regarding when a glueball field corresponding to a given branch cut
should be set to zero before extremizing the off-shell glueball superpotential.
The discussion of IKRSV was restricted to $\CN=1$ gauge theories with an
adjoint and no flavors, so the generalization to the case with
fundamental flavors is obviously the next task.

Our second aim of this paper is to carry out this task. The main result
is the following.  Assuming $N_f$ poles around a cut associated with
gauge group $U(N_{c,i})$, when $N_f\geq N_{c,i}$ there are situations in
which we should set $S_i=0$ in matrix model computations.  More
concretely, situations with $S_i=0$ belong to either of the following
two branches: the baryonic branch for $N_{c,i}\leq N_f< 2N_{c,i}$, or
the $r=N_{c,i}$ non-baryonic branch for $N_f\ge 2N_{c,i}$. Moreover,
when $S_i=0$, the gauge group is completely broken and there should
exist some extra, charged massless field which is not incorporated in
matrix model.

In section \ref{sec:bg}, as background, we review basic materials for
${\cal N}=1$ supersymmetric $U(N_c)$ gauge theory with an adjoint chiral
superfield, and $N_f$ flavors of quarks and anti-quarks.  The chiral
operators and the exact effective glueball superpotential are given. We
study the vacuum structure of the gauge theory at classical and quantum
levels.  We review also the main results of IKRSV.\ \ In addition to all
these reviews, we present our main motivations of this paper.


In section \ref{sec:1cut}, we apply the formula for the off-shell
superpotential obtained in \cite{CSW-II} to the case with quadratic tree
level superpotential, and solve the equation of motion derived from it.
We consider what happens if one moves $N_f$ poles associated with
flavors on the second sheet through the cut onto the first sheet,
on-shell.
Also, in subsection \ref{subsec:general'n_IKRSV}, we briefly touch the
matter of generalizing IKRSV in the one cut model.

In section \ref{sec:2cut}, we consider cubic tree level superpotential.
On the gauge theory side, the factorization of the Seiberg--Witten curve
provides an exact superpotential.  We reproduce this superpotential by
matrix model, by extremizing the effective glueball superpotential with
respect to glueball fields after setting the glueball field to zero when
necessary.  We present explicit results for $U(3)$ theory with all
possible breaking patterns and different number of flavors
($N_f=1,2,3,4$, and 5).

In section \ref{sec:conclusion}, after giving concluding remarks, we
repeat the procedure we did in previous sections for $SO(N_c)/USp(2N_c)$
theories, briefly.

In the appendix, we present some proofs and detailed calculations which
are necessary for the analysis in section \ref{sec:2cut}.

Since string theory results in the dual Calabi--Yau geometry are
equivalent to the matrix model results, we refer to them synonymously
through the paper.  There exist many related works to the present paper.
For a list of references, we refer the reader to \cite{Argurio:2003ym}.

\section{Background}
\setcounter{equation}{0}
\label{sec:bg}

\indent

In this section, we will summarize the relevant background needed for
the study of ${\cal N}=1$ supersymmetric gauge theory with matter
fields.

\subsection{The general picture of matrix model with flavors}

\indent

The generalized Konishi anomaly interpretation to the matrix model
approach for $\CN=1$ supersymmetric gauge theory with flavors was given
in \cite{seiberg02,CSW-II}. Here we make only a brief summary on some
points we will need.

Let us consider $\CN=1$ supersymmetric $U(N_c)$ gauge theory, coupled to
an adjoint chiral superfield $\Phi$, $N_f$ fundamentals $Q^f$, and $N_f$
anti-fundamentals $\Qt_\ft$.  The tree level superpotential is taken to
be
\bea 
\label{csw2-2.1}
W_{\rm tree} & = & \Tr\, W(\Phi)+\sum_{f,\ft}\W Q_{\W f}\, m^{\W f}{}_{f}(\Phi) Q^f~,
\eea
where the function $W(z)$ and the matrix $m_f^\ft(z)$ 
are polynomials 
\bea 
W(z) & = & \sum_{k=0}^{n} \frac{g_k z^{k+1}}{k+1}, \qquad
m^{\W f}{}_{f}(z)  =  \sum_{k=1}^{l+1} (m_k)^{\W f}{}_{f} z^{k-1}~.
\nonu
\eea
Classically we can have the ``pseudo-confining vacua'' where
the vacuum expectation values of  $Q$, $\W Q$  are zero, or the
``Higgs vacua'' where the vacuum expectation values of $Q$, $\W Q$ are 
nonzero so that the total rank of the remaining gauge groups is
reduced. 
These two vacua, which seem to have a big difference classically, are
not fundamentally distinguishable from each other in the quantum theory
and in fact can be continuously transformed into each other, as we will
review shortly, in the presence of flavors \cite{CSW-II}.

Supersymmetric vacua of gauge theory are characterized by the vacuum
expectation values of chiral operators \cite{Svrcek:2003az}.  They are
nicely packaged into the following functions called resolvents
\cite{CSW-II,seiberg02}: \footnote{We set 
$w_\alpha(z)\equiv \frac{1}{
4\pi}\Ev{\Tr \frac{W_\a}{ z-\Phi}}$ to zero because in supersymmetric
vacua $w_\alpha(z)=0$.}
\begin{align}
 T(z) & =  \Ev{\Tr \frac{1}{ z-\Phi}}, \label{csw2-2.10-a}\\
 R(z) & =  -\frac{1}{ 32\pi^2} \Ev{\Tr 
\frac{W_\a W^\a}{ z-\Phi}}, 
\label{csw2-2.10-c}
\\
 M(z)^f{}_{\W f} & =  \Ev{\W Q_{\W f} \frac{1}{ z-\Phi} Q^f} 
\label{csw2-2.10-d}
\end{align}
where $W_\a$ is (the lowest component of) the field strength superfield.
Classically, $R(z)$ vanishes while $T(z)$, $M(z)$ have simple poles on
the complex $z$-plane at infinity and at the eigenvalues of $\Phi$.
Each eigenvalue of $\Phi$ is equal to one of zeros of $W'(z)$ or $B(z)$,
where
\bea  
\label{csw2-2.2-1}
W'(z)= g_n \prod_{i=1}^n (z-a_i),\qquad
B(z) \equiv \det m(z)=B_L \prod_{I=1}^L (z-z_I).
\eea
In the pseudo-confining vacuum, every eigenvalue of $\Phi$ is equal to
$a_i$ for some $i$.  On the other hand, in the Higgs vacuum, some
eigenvalues of $\Phi$ are equal to $z_I$ for some $I$.

In the quantum theory, the resolvents
\eqref{csw2-2.10-a}, \eqref{csw2-2.10-c}
and \eqref{csw2-2.10-d} are determined by the
generalized Konishi anomaly equations \cite{csw1, seiberg02,CSW-II}:
\begin{equation}
\begin{split}
%
~[W'(z) T(z)]_-+ \Tr [m'(z) M(z)]_-  & =  2 R(z) T(z),  \\
~[W'(z) R(z)]_- & =  R(z)^2, \\
~[(M(z) m(z))^{f'}_{~f}]_- & =  R(z) \delta^{f'}_{f}, \\
~[( m(z) M(z))^{\W f'}_{~ \W f}]_- & =  R(z) \delta^{\W f'}_{\W f},
\end{split}
\label{csw2-2.14} 
\end{equation}
where the notation $[~]_-$ means to drop the nonnegative powers in a
Laurent expansion in $z$.   From the second 
equation of (\ref{csw2-2.14}),
one obtains \cite{CDSW}
\bea 
R(z) = \frac{1}{ 2} \left( W'(z)-\sqrt{W'(z)^2+f(z)}\right),
\nonu
\eea 
where $f(z)$ is a polynomial of degree $(n-1)$ in $z$.  This implies
that in the quantum theory the zeros $z=a_i$ ($i=1,2,\dots, n$) of
$W'(z)$ are blown up into cuts $A_i$ along
intervals\footnote{\label{phys_mean_cut} $a_i^{\pm}$ are generally
complex and in such cases we take $A_i$ to be a straight line connecting
$a_i^-$ and $a_i^+$.  Note that there is no physical meaning to the
choice of the cut; it can be any path connecting $a_i^-$ and $a_i^+$.}
$[a_i^-,a_i^+]$ by the quantum effect represented by $f(z)$, and the
resolvents \eqref{csw2-2.10-a}--\eqref{csw2-2.10-d} are defined on a
double cover of the complex $z$-plane branched at the roots $a_i^\pm$ of
$ W'(z)^2+f(z)$.  This double cover of the $z$-plane can be thought of
as a Riemann surface $\Sigma$ described by the matrix model curve
\bea
\label{csw2-2.18}
\Sigma: \quad y_m^2= W'(z)^2+f(z).
\eea
This curve is closely related to the factorization form of 
$\CN=2$ curve in the strong coupling analysis.

Every point $z$ on the $z$-plane is lifted to two points on the
Riemann surface $\Sigma$ which we denote by $q$ and $\qt$ respectively.
For example, $z_I$ is lifted to $q_I$ on the first sheet and $\W q_I$ on
the second sheet.  We write the projection from $\Sigma$ to the
$z$-plane as $z_I=z(q_I)=z(\widetilde q_I)$, following the notation of
\cite{CSW-II}.
 
The classical singularities of the resolvents $T(z)$, $M(z)$ are
modified in the quantum theory to the singularities on $\Sigma$, as
follows.
For $T(z)$, the classical poles at
$z_I$ are lifted to poles at $q_I$ or $\W q_I$, depending on which
vacuum the theory is in, while the classical poles at $a_i$ with residue
$N_{c,i}$ are replaced by cuts with periods $\frac{1}{ 2\pi i}\oint_{A_i}
T(z) dz=N_{c,i}$.  For $M(z)$, the classical poles at $z_I$ are also lifted
to poles at $q_I$ or $\W q_I$.
More specifically, by solving the last two equations of (\ref{csw2-2.14}),
one can show \cite{CSW-II}
\begin{align}
\label{Higgs-M} 
M(z) = R(z) \frac{1}{m(z)}
 -\sum_{I=1}^L \frac{ (1-r_I)R(q_I) }{ (z-z_I)} \frac{1}{ 2\pi i}\oint_{q_I} \frac{dx}{ m(x)} 
 -\sum_{I=1}^L \frac{ r_I R(\W q_I)}{(z-z_I)} \frac{1}{ 2\pi i} \oint_{\W q_I} \frac{dx}{ m(x)},
\end{align}
where $(q_I,\W q_I)$ are the lift of $z_I$ to the first sheet and to the
second sheet of $\Sigma$, and $r_I=0$ for  poles on the second sheet
and $r_I=1$ for  poles on the first sheet.
Furthermore, for $T(z)$, by solving the first equation of
(\ref{csw2-2.14}),
\bea \label{Higgs-T-z} T(z) & = & \frac{ B'(z)}{ 2
B(z)}- \sum_{I=1}^L \frac{ (1-2 r_I) y(q_I)}{ 2y(z) (z-z_I)}+
\frac{c(z)}{ y(z)},
\eea
where
\bea \label{c-z} c(z) & = & \braket{ \Tr
\frac{W'(z)-W'(\Phi) }{ z-\Phi}} -\frac{1}{ 2} \sum_{I=1}^{L} \frac{
W'(z)-W'(z_I)}{ z-z_I}. 
\eea
Practically it is hard to use (\ref{c-z}) to obtain $c(z)$ and we use
the following condition instead:
\bea
\label{off-shell} \frac{1}{ 2\pi i}\oint_{A_i} T(z) dz=N_{c,i}. 
\eea 
Finally, the exact, effective glueball superpotential is given by \cite{CSW-II}
\begin{multline}
 \label{Wexact-y} 
 W_{\rm eff}  =  
 -\frac{1}{ 2} \sum_{i=1}^n N_{c,i} \int_{\widehat{B}^r_i} y(z) dz
 -\frac{1}{ 2} \sum_{I=1}^L (1-r_I)\int_{\widetilde{q}_I} ^{\widetilde{\Lambda}_0} y(z) dz
 -\frac{1}{ 2} \sum_{I=1}^L r_I \int_{q_I} ^{\widetilde{\Lambda}_0} y(z) dz 
 \\
 +\frac{1}{ 2}(2N_c-L) W(\Lambda_0) 
 +\frac{1}{ 2} \sum_{I=1}^L W(z_I) 
 -\pi i(2N_c-L)S+ 2\pi i \tau_0 S +2\pi i \sum_{i=1}^{n-1} b_i S_i,
\end{multline}
where
\bea \label{scale} 
2\pi i \tau_0 =\log\left(
\frac{B_L \Lambda^{2N_c-N_f} }{ \Lambda_0^{2N_c-L}} \right).
\eea
Here, $\Lambda_0$ is the cut-off of the contour integrals, $\Lambda$ is
the dynamical scale, $S\equiv \sum_{i=1}^n S_i$, and $b_i\in
\mathbb{Z}$.  $\widehat{B}_i^r$ is the regularized contour from
$\widetilde{\La}_0$ to $\La_0$ through the $i$-th cut and $\La_0$ and
$\widetilde{\La}_0$ are the points on the first sheet and on the second
sheet, respectively.
The glueball field is defined as 
\bea 
S_i =\frac{1}{2\pi i} \oint_{A_i} R(z) dz.
\nonu
\eea

In the above general solutions \eqref{Higgs-M}, \eqref{Higgs-T-z}, we
have $r_I=1$ or $r_I=0$, depending on whether the pole is on the first
sheet or on the second sheet of $\Sigma$, respectively.  The relation
between these choices of $r_I$ and the phase of the system is as
follows.  Let us start with all $r_I=0$, i.e., all the poles are on the
second sheet. This choice corresponds to the pseudo-confining vacua
where the gauge group is broken as $U(N_c)\to \prod_{i=1}^{n} U(N_{c,i})$
with $\sum_{i=1}^{n} N_{c,i}=N_c$. Now let us move a single pole through,
for example, the $n$-th cut to the first sheet. This will break the
gauge group as $\prod_{i=1}^{n} U(N_{c,i})\to \prod_{i=1}^{n-1} U(N_{c,i})
\times U(N_{c,n}-1)$.  
Note that the rank of the last factor is now $(N_{c,n}-1)$ so
that $\sum_{i=1}^{n} N_{c,i}=N_c-1<N_c$.  Namely, the gauge group is Higgsed
down. In this way, by passing poles through cuts, one can go
continuously from the pseudo-confining phase to the Higgs phase, as
advocated before.

However, if we consider this process of passing poles through a cut to
the first sheet {\em on-shell\/}, then there should be an obstacle at a
certain point. For example, if initially we have $N_{c,n}=1$, 
after passing
a pole we would end up with an $U(0)$.  This sudden jump of the number
of $U(1)$'s in the low energy gauge theory is not a smooth physical
process, because the number of massless particles (photons) changes
discontinuously.  So we expect some modifications to the above
picture. In \cite{CSW-II}, it was suggested that in an on-shell process,
the $n$-th cut will close up in such a situation so that the pole cannot
pass through. It is one of our motivations to show that this is indeed
true.  More precisely, the cut does not close up completely and the pole
can go through a little bit further to the first sheet and then will be
bounced back to the second sheet.

\subsection{The vacuum structure}

\indent

In the last subsection we saw that different distributions of poles over
the first and the second sheets correspond to different phases of the
theory.  In this subsection we will try to understand this vacuum
structure of the gauge theory at both classical and quantum levels for a
specific model (for more details, see 
\cite{APS,ckm,ckkm,bfhn,afo1,afo2}).
For simplicity we will focus on $U(N_c)$ theory with $N_f$ flavors and
the following tree level superpotential \footnote{We used the convention
of \cite{DJ} for the normalization of the second term. Different choices
are related to each other by redefinition of $\W Q$ and $Q$.}
%
%
%
%
\bea \label{quadraticpot}
W_{\rm tree} & = & \frac{1}{2} m_A \Tr\,\Phi^2
-
\sum_{I=1}^{N_f} \widetilde{Q}_{I}
 (\Phi + m_f) Q^{I}.
\eea
%
This corresponds to taking polynomials in \eqref{csw2-2.1} as
\begin{align}
 W(z)&=\frac{m_A}{ 2}z^2,\qquad 
 m^{\widetilde I}_{~I}(z)=-(z+m_f)\delta_{I}^{\widetilde I}.
\nonu
\end{align}
All $N_f$ flavors have the same mass $m_f$, and the mass function
defined in \eqref{csw2-2.2-1} is given by
\begin{align}
 B(z)&=(-1)^{N_f}(z+m_f)^{N_f}.
\nonu
\end{align}
Therefore, poles associated with  flavors are located at
\begin{align}
 z_I=-m_f\equiv z_f, \qquad I=1,2,\dots, N_f~.
 \label{def_zf}
\end{align}
In the quantum theory, some of these poles are lifted to $q_f$ on the
first sheet and others are lifted to $\qt_f$ on the second sheet.


The $D$- and $F$-flatness for the superpotential (\ref{quadraticpot})
is given by
\begin{align*}
0  =  [ \Phi, \Phi^{\dagger}],   \quad 
0  =  Q Q^{\dagger}- \W Q^{\dagger} \W Q, \quad
0  =  m_A \Phi- Q \W Q , \quad
0  =  (\Phi + m_f) Q= \W Q  (\Phi + m_f).
\end{align*}
Solutions are a little different for $m_f\neq 0$ and $m_f=0$,
because $m_f=0$ is the root of $W'(z)=z$. The case of $W'(-m_f)=0$
was discussed in \cite{APS,bfhn} which we will refer to as the
classically massless case.

In the $m_f\neq 0$ case, the solution is given by
\begin{equation}\label{sol2DF}
\begin{split}
\Phi & =  \left[ \begin{array}{cc}  -m_f I_{K\times K} & 0 \\
0 & 0_{(N_c-K)\times (N_c-K)} \end{array} \right], \\
Q   & =   \left[ \begin{array}{cc}  A_{K\times K} & 0 \\
0 & 0_{(N_c-K)\times (N_f-K)} \end{array} \right], ~~~
 ^t \W Q  =  \left[ \begin{array}{cc}  ^t\! \W A_{K\times K} &
\W B_{K\times (N_f-K)} \\
0 & 0_{(N_c-K)\times (N_f-K)} \end{array} \right]
\end{split}
\end{equation}
with 
\bean
- m_f m_A I_{K\times K}= A \,^t\!\W A, ~~~~~~
A A^{\dagger}= \W A^{\dagger} \W A+ \W B^* \,^t\!\W B.
\eean
The gauge group is Higgsed down to $U(N_c-K)$ where 
\bea 
K_{m_f\neq 0} \leq \min\left(N_c, N_f\right).
\nonu
\eea
To understand the range of $K_{m_f\neq 0} $, first note that the
$\Phi$ breaks the gauge group as $U(N_c)\to U(K)\times U(N_c-K)$. Now
the $U(K)$ factor has effectively $N_f$ massless flavors and because
$\ev{\W Q Q} \neq 0$, $U(K)$ is further Higgsed down to 
$U(0)$.

For $m_f=0$, we have $\Phi=0$ and $Q, \W Q$ are still of the above form
\eqref{sol2DF} with one special requirement: $\W A=0$. Because of this
we have
\bea 
K_{m_f=0} \leq \min\left(N_c, \left[\frac{N_f}{ 2}\right]\right),
\nonu
\eea
where $[~]$ means the integer part.  The integer $K_{m_f=0}$ precisely
corresponds to the $r$-th branch discussed in \cite{APS,bfhn}.  The
$m_f=0$ case is different from the $m_f\neq 0$ case as follows. First,
$\Phi$ does not break the $U(N_c)$ gauge group, i.e., $U(N_c)\to
U(N_c)$. Secondly, the $r$-th branch is the intersection of the Coulomb
branch in which $\ev{\W Q Q}=0$ and the Higgs branch in which $\ev{\W Q
Q}\neq 0$, whereas for $m_f\neq 0$ the vacuum expectation value $\ev{\W Q Q}$ must be nonzero
and the gauge group must be Higgsed down. For these reasons, $K_{m_f\neq
0}$ and $K_{m_f=0}$ have different ranges.

The above classical classification of  $r$-th branches is also valid
in the quantum theory (including the baryonic branch). 

The quantum $r$-th branch can also be discussed by using the
Seiberg--Witten curve.  In the $r$-th branch, the curve factorizes as
\begin{align*}
 y_{\CN=2}^2
 &= P_{N_c}(x)^2 -4 \Lambda^{2N_c-N_f} (x+m_f)^{N_f}\\
 &=(x+m_f)^{2r} \left[ P_{N_c-r}^2(x) - 4 \La^{2N_c-N_f} (x+m_f)^{N_f-2r} \right].
\end{align*}
Because $N_c-r \geq 0$ (coming from $P_{N_c-r}(x)$) and $N_f -2r \geq 0$
(coming from the last term), we have $r \leq N_c$ and 
$r \leq N_f/2 $,
which leads to the range
\bea 
r \leq \min\left(N_c, \left[\frac{N_f}{ 2}\right]\right).
\label{range_r-vac}
\eea
The relation between this classification of the Seiberg--Witten curve
and the above classification of  
$r$-branches, in the $m_f\neq 0$ and
$m_f=0$ cases, is as follows.  In the $m_f=0$ case, we have one-to-one
correspondence where the $r$ is identified with $K_{m_f=0}$. In the
$m_f\neq 0$ case, for a given $r$ of the curve, there exist two cases:
either $K_{m_f\neq 0}=r$ for $K_{m_f\neq 0}\leq [N_f/2]$, or $K_{m_f\neq
0}=N_f-r$ for $K_{m_f\neq 0}\geq [N_f/2]$.

\subsection{The work of IKRSV}

\indent

Now we discuss another aspect of the model. In the above, we saw that
there is a period condition (\ref{off-shell}) for $T(z)$.  So, if for
the $i$-th cut we have $\oint_{A_i} T(z) dz=
N_{c,i}=0$, then it seems that,
in the string theory realization of the gauge theory, there is no RR
flux provided by D5-branes through this cut and the cut is
closed. Because of this, it seems that we should set the corresponding
glueball field $S_i=0$.  Based on this naive expectation, Ref.\
\cite{Kraus:2003jf} calculated the effective superpotential of
$USp(2N_c)$ theory with an 
antisymmetric tensor by the matrix model, which
turned out to be different from the known results obtained by holomorphy
and symmetry arguments (later Refs. \cite{krs,ac} confirmed this
discrepancy).

This puzzle intrigued several papers \cite{agaetal, cach, IKRSV, afo2,
Matone:2003bx, ll, Naculich:2003ka, Gomez-Reino:2004rd}.  In particular,
in \cite{cach}, it was found that although $N_{c,i}=0$, we cannot set
$S_i=0$. The reason became clear by later studies. Whether a cut closes
or not is related to the total RR flux which comes from both D5-branes
and orientifolds. For $USp(2N_c)$ theory with antisymmetric tensor,
although the RR flux from D5-branes is zero, there exists RR flux coming
from the orientifold with positive RR charges, thus the cut does not
close. That the cut does not close can also be observed from the
Seiberg--Witten curve \cite{afo2} where for such a cut, we have two
single roots in the curve, instead of a double root.  All these results
were integrated in \cite{IKRSV} for $\CN=1$ gauge theory with
adjoint. Let us define $\WH N_c=N_c$ for $U(N_c)$, $N_c/2-1$ for
$SO(N_c)$ and $2N_c+2$ for $USp(2N_c)$. Then the conclusion of
\cite{IKRSV} can be stated as
\begin{quote}
 If $\WH N_c>0$, we should include $S_i$ and extremize $W_{\rm
 eff}(S_i)$ with respect to it. On the other hand, if $\WH N_c\leq 0$, 
 we just set $S_i= 0$ instead.
\end{quote}
In \cite{IKRSV}, it was argued that this prescription of setting $S_i=0$
can be explained in string theory realization by considering an extra
degree of freedom which corresponds to the D3-brane wrapping the blown
up $S^3$ and becomes massless in the $S\to 0$ limit
\cite{Strominger:1995cz}.  Our second motivation of this paper is to
generalize this conclusion to the case with flavors.
We will discuss the precise condition when one should set $S_i=0$ in
order to get agreement with the gauge theory result, in the case with
flavors.

\subsection{Prospects from the strong coupling analysis}
\label{subsec:prospects_strong_cpl}

\indent

Before delving into detailed calculations, let us try to get some
general pictures from the viewpoint of factorization of the
Seiberg--Witten curve.  
Since we hope to  generalize IKRSV, we are interested in the case
where some $S_i$ vanish.  Because $S_i$ is related to the size of a
cut in the matrix model curve, which is essentially the same as the
Seiberg--Witten curve, we want some cuts to be closed in the
Seiberg--Witten curve.  Namely, we want a double root in the
factorization of the curve, instead of two single roots.

%

For $U(N_c)$ theory with $N_f$ flavors of the same mass $m_f=-z_f$, tree
level superpotential \eqref{csw2-2.1}, 
and breaking pattern
$U(N_c)\to \prod_{i=1}^n U(N_{c,i})$, the factorization form of the
curve is \cite{bfhn}
\begin{equation}
  \label{factor-UN}
\begin{split}
 P_{N_c}(z)^2-4\La^{2N_c-N_f} (z-z_f)^{N_f} &= H_{N-n}(z)^2 F_{2n}(z),\\
 F_{2n}(z) &= W'(z)^2+f_{n-1}(z),
\end{split}
\end{equation}
where the degree $2n$ polynomial $F_{2n}(z)= W'(z)^2+f_{n-1}(z)$
generically has $2n$ single roots.  How can we have a double root
instead of two single roots?

For a given fixed mass, for example $z_f=a_1$, there are three cases where
we have a double root, as follows.  (a) There is no $U(N_{c,1})$ 
group factor
associated with the root $a_1$, namely $N_{c,1}=0$. (b) The $U(N_{c,1})$
factor is in the baryonic branch. This can happen for $N_{c,1}\leq N_f<
2N_{c,1}$. (c) The $U(N_{c,1})$ factor is in the $r$-th non-baryonic
branch with $r=N_{c,1}$. This can happen only for $N_f\geq
2N_{c,1}$. Among these three cases, (b) and (c) \cite{bfhn} are new for
theories with flavors, and will be the focus of this paper.
However, it is worth pointing
out that the factorization form in the
cases (b) and (c) are not the one given in (\ref{factor-UN}) for a fixed
mass, but the one given in (\ref{Nocorrection}). 

Can we keep the factorization form (\ref{factor-UN}) while having an
extra double root? We can, but instead of a fixed mass we must let the
mass  ``floating,'' which means the following. There will be multiple
solutions to the factorization form (\ref{factor-UN}), and for any given
solution the $2n$ single roots of $F_{2n}(x)$, denoted by $a_i^\pm$,
$i=1,\dots,n$, are functions of $z_f$.  Now, we tune $z_f$ so that
$a_i^+(z_f)=a_i^-(z_f)$, i.e., so that two single roots combine into one
double root. Since for different solutions this procedure will lead to
different values of $z_f$, we call this situation the ``floating'' mass.

Now we have two ways to obtain extra double roots: one is with a
fixed mass, but to go to the baryonic or the $r=N_{c,1}$ 
non-baryonic branch,
while the other is to start with a general non-baryonic branch but using
a floating mass. In fact it can be shown that these two methods are
equivalent to each other when the double root is produced.
In the calculations in section \ref{sec:2cut} we will use the floating
mass to check our proposal.

\section{One cut model---quadratic tree level superpotential }
\setcounter{equation}{0}
\label{sec:1cut}

\indent

In this section we will study whether a cut closes up if one tries to
pass too many poles through it.  If the poles are near the cut, the
precise form of the tree level superpotential (namely the polynomials
$W(z)$, $m^{\tilde{f}}{}_f(z)$) is inessential and we can simplify the
problem to the quadratic tree level superpotential given by
(\ref{quadraticpot}).
For this superpotential, we will compute the effective glueball
superpotential using the formalism reviewed in the previous section.
Then, by solving the equation of motion, we study the {\em on-shell\/}
process of sending poles through the cut, 
and see whether the poles can pass  or not.

Also, on the way, we make an observation on the relation between the
exact superpotential and the vacuum expectation value of the tree level
superpotential.

\subsection{The off-shell $W_{\rm eff}$, $M(z)$ and $T(z)$}

\indent

First, let us compute the effective glueball superpotential for the
quadratic superpotential \eqref{quadraticpot}.  The matrix model curve
\eqref{csw2-2.18} is related to $W'(z)$ in this case as
\bea \label{matrix-curve}
y_m^2 & = & W'(z)^2+f_{0}(z)= m_A^2 z^2-\mu \equiv
m_A^2 (z^2-\W \mu),~~~~~~~\W \mu=\frac{\mu }{ m_A^2}.
\eea

Let us consider the case with $K$ poles on the first sheet at $q_I=q_f$,
for which $r_I=1$, and with $(N_f-K)$ poles on the second sheet at $\W
q_I=\W q_f$, for which $r_I=0$ (recall that $(q_f,\W q_f)$ is the lift
of $z_f$
defined in \eqref{def_zf}).
Using the curve (\ref{matrix-curve}) and various formulas summarized in
the previous section, one can compute
\bea
S & = &  \frac{1}{ 2\pi i}\oint_{A} R(z) dz=
\frac{m_A\W \mu }{ 4} =\frac{  \mu}{ 4 m_A},
\nonu \\
\Pi & = &2\int_{\sqrt{\W \mu}}^{\Lambda_0} y(z) dz
 =   m_A\Lambda_0^2 - 2 S - 2 S \log \frac{ \Lambda_0^2 m_A }{ S},\notag \\
\Pi_{f,I}^{r_I=0} 
& = & \int_{\W q_I}^{\W \Lambda_0} y(z)d z
 = -\int_{q_I}^{\Lambda_0} y(z)d z\notag\\
&=& \frac{- m_A\Lambda_0^2}{ 2}- 2S \log \frac{ z_I}{ \Lambda_0}
+2 S\left[ -\log\left( \frac{1}{ 2}+ \frac{1}{ 2}\sqrt{ 1- 
\frac{4 S}{ m_A z_I^2}}\right)
+\frac{m_A z_I^2 }{ 4 S}\sqrt{ 1- \frac{4 S}{ m_A z_I^2}}+
\frac{1}{ 2}\right],
\notag\\
\Pi_{f,I}^{r_I=1} & = & \int_{q_I}^{\W\Lambda_0} y(z) dz
=-\int_{\W q_I}^{\Lambda_0} y(z) dz \notag\\
&=& \frac{- m_A\Lambda_0^2}{ 2}
- 2S \log \frac{ z_I}{ \Lambda_0}
+2 S\left[ -\log\left( \frac{1}{ 2}- 
\frac{1}{ 2}\sqrt{ 1- \frac{4S}{ m_A z_I^2}}\right)
-\frac{m_A z_I^2 }{ 4 S}
\sqrt{ 1- \frac{4S}{ m_A z_I^2}}+\frac{1}{ 2}\right],
\notag
\eea
where we dropped $\CO( {1/ \Lambda_0})$ terms.  
We have traded $q_I$, $\W q_I$ for $z_I$ in the square roots,
so that the sign convention is such that $\sqrt{1- \frac{4 S}{ m_A
z_I^2}}\sim 1-\frac{2 S}{ m_A z_I^2}$ and $\sqrt{z_I^2- 
\frac{4 S}{ m_A
}}\sim z_I$ for $|z_I|$ very large. 
Substituting this into (\ref{Wexact-y}), we obtain
\bea 
W_{\rm eff}(S) & = &S \left[ N_c+ \log \left( \frac{m_A^{N_c}
\Lambda^{2N_c-N_f} \prod_I z_I }{ S^{N_c}} \right) \right]\nonumber\\
& & -\sum_{I,r_I=0} S
\left[ -\log\left( \frac{1}{ 2}+ \frac{1}{ 2}
\sqrt{ 1- \frac{4S}{ m_A z_I^2}}
\right)
+\frac{m_A z_I ^2 }{ 4 S}
\left(\sqrt{ 1- \frac{4 S}{ m_A z_I ^2}} -1\right)+
\frac{1}{ 2}\right]
\nonumber \\
& &-\sum_{I,r_I=1} S
\left[ -\log\left( \frac{1}{ 2}- \frac{1}{ 2}
\sqrt{ 1- \frac{4 S}{ m_A z_I^2}}
\right)
+\frac{m_A z_I ^2}{ 4 S}
\left(-\sqrt{ 1- \frac{4S}{ m_A z_I ^2}} -1\right)+
\frac{1}{ 2}\right],~~~~~~
 \label{quadratic-exact}
\eea
where $\Lambda$ is the dynamical scale of the corresponding ${\cal N}=2$
gauge theory defined in \eqref{scale}.

Let us compute resolvents also.
The resolvent $M(z)$ is an $N_f\times N_f$ matrix.  Using
(\ref{Higgs-M}), we find that for the $I$-th eigenvalue
\bea 
M_{I}(z) & = & \frac{ -R(z) }{  (z- q_I)}
+\frac{(1-r_I) R(q_I) }{ (z- q_I)}+
\frac{r_I R(\W q_I) }{  (z- q_I)}.
\nonu
\eea
Expanding $M_I(z)$ around $z=\infty$ we can read off the
following vacuum expectation values
\begin{gather*}
 \ev{\W Q Q}_{I} 
    =  \frac{m_A}{ 2} 
{ \Bigl[z_I +(2 r_I-1)\sqrt{z_I^2-\W \mu}\Bigr]}, ~~
 \ev{ \W Q \Phi Q}_{I} 
    = \frac{m_A}{ 4} 
{ \Bigl[ 2 z_I^2 +2(2r_I-1) z_I\sqrt{z_I^2-\W \mu}-\W \mu\Bigr]},\\
 \Ev{\sum_I -( \W Q \Phi Q- z_I \W Q Q)}  =  
\frac{ N_f m_A \W \mu}{ 4} = N_f S.
\end{gather*}
There is something worth noting here. One might naively expect that the
exact superpotential is simply the vacuum expectation value of the tree
level superpotential (\ref{quadraticpot}), as is the case without
flavors.  However, this naive expectation is wrong!  Although we have
$\ev{W_{\rm tree,fund}} = \ev{-\sum_I ( \W Q \Phi Q- z_I \W Q Q)}\neq 0$
if $S\neq 0$, we still have
\bea \label{promise} 
W_{\text{eff,on-shell}} = \braket{ \frac{m_A}{ 2} \Tr\, \Phi^2}~,
\eea 
as we will see shortly. The reason for (\ref{promise}) can be explained
by symmetry arguments \cite{IS-1,Eli}. Although the tree level part
$\ev{W_{\rm tree,fund}}$ for fundamentals is generically nonzero and
also contributes to $W_{\rm low}$, the contribution is precisely
canceled by the dynamically generated superpotential $W_{\rm dyn}$,
leaving only the $\Phi$ part of $W_{\rm tree}$
\footnote{We would like
to thank K. Intriligator for explaining this point to us.}.

Let us calculate the resolvent $T(z)$ also.  For the present example,
$c(z)=m_A(N_c-\frac{N_f}{ 2})$ from (\ref{c-z}).  Therefore, using
(\ref{Higgs-T-z}), we obtain the expansion of the resolvent $T(z)$:
%
\bea
T(z) & = & \frac{N_c}{ z}+\sum_{I}\left[ \frac{ z_I}{ 2}+
\frac{ 2 r_I-1}{ 2} \sqrt{ z_I^2- \W \mu} \right] \frac{1}{ z^2} \nonumber \\
& & + \left[ \frac{ \W \mu(2N_c-N_f) }{ 4}+ \sum_{I}\left(
\frac{ z_I^2}{ 2}+ \frac{2 r_I-1}{ 2} z_I  
\sqrt{ z_I^2- \W \mu} \right)
\right]\frac{1}{ z^3}+\cdots 
 \label{exact-T(z)}
\eea
From this we can read off $\braket{\Tr\,\Phi^n}$.  For example, for
$K=0$ we have
\bea 
\braket{\Tr\, \Phi} =  \frac{N_f}{ 2} 
\left(z_f-\sqrt{ z_f^2-\W \mu}\right) , \qquad
\braket{\Tr\, \Phi^2}  =   \frac{N_f}{ 2} 
z_f\left(z_f-\sqrt{ z_f^2-\W \mu}\right)    + 
\frac{(2N_c-N_f) \W \mu }{ 4}.
   \nonumber 
\nonu
\eea
%

\subsection{The on-shell solution}

\indent

Now we can use the above off-shell expressions to find the on-shell
solution. First we rewrite the superpotential as
\bea
W_{\rm eff} & = &S \left[ N_c+ \log \left(\frac{
\Lambda_1^{3N_c-N_f}}{ S^{N_c} }\right) \right]\notag\\
& & - (N_f-K)S
\left[ -\log\left( \frac{z_f}{ 2}+ \frac{1}{ 2}
\sqrt{ z_f^2- \frac{4S}{ m_A }}
\right)
+\frac{m_A z_f}{ 4 S}
\left(\sqrt{ z_f^2- \frac{4 S}{ m_A}} -z_f\right)+
\frac{1}{ 2}\right]\notag\\
& &-K S \left[ -\log\left( \frac{z_f}{ 2}- 
\frac{1}{ 2}\sqrt{ z_f^2- \frac{4 S}{
m_A }} \right) +\frac{m_A z_f }{ 4 S} \left(-\sqrt{ z_f^2- 
\frac{4S}{
m_A }} -z_f\right)+\frac{1}{ 2}\right],
\label{superpot_quad}
\eea
where $\Lambda_1^{3N_c-N_f}\equiv m_A^{N_c} \Lambda^{2N_c-N_f}$.  We
have set all $z_I$ to be $z_f$, i.e., all masses are the same, as in
\eqref{def_zf}.  From this we obtain 
\footnote{
It
is easy to check that the equation (\ref{equ-W-1}) with parameters
$(N_c,N_f,K)$ is the same as the one with parameters $
(N_c-r,N_f-2r,K-r)$.  Also from the expression (\ref{exact-T(z)}) it
is straightforward to see we
have $\braket{\Tr \Phi^2}_{N_c,N_f,K}=\braket{\Tr
\Phi^2}_{N_c-r,N_f-2r,K-r} +r z_I^2$.  
All of these facts are the result
of the ``addition map'' observed in \cite{bfhn}. Furthermore, one can
show that both (\ref{equ-W-1}) and (\ref{exact-T(z)}) for $K=0$ are
exactly the same as the one given by the strong coupling analysis in
\cite{DJ} and the weak coupling analysis in \cite{De, EHU}.
}
\bea 
\label{equ-W-1}
0  =   
    \log  \biggl(\!\frac{\Lambda_1^{3N_c-N_f}}{ S^{N_c}}\!\biggr)
 + K \log 
\left( \frac{ z_f- \!\sqrt{ z_f^2- \frac{4 S}{ m_A }}  }{ 2} \right)
 +(N_f-K) \log \left( 
\frac{ z_f+ \!\sqrt{ z_f^2- \frac{4 S}{ m_A }}  }{ 2} \right),
\eea
or
\footnote{As we mentioned before, for $m_f\neq 0$ the allowed Higgs branch
requires $K\leq N_f$ but the strong coupling analysis gives
$K\leq N_f/2$. The resolution for that puzzle is that
if $K> N_f/2$, it is given by $(N_f-K)$-th branch of the curve.
Since the same $r$-th branch of the curve gives both
$r$ and $(N_f-r)$ Higgs branches, we expect that
$r$ and $(N_f-r)$ Higgs branches are  related. This 
relation is given by $\W S= S b^2$, $\W z_f= z_f b$ with
$b = \frac{m_A \Lambda^2 }{ S}$. It can be shown that with the above 
relation, the equation of motion of $S$ for the 
$K$-th branch is changed to the equation of motion of $\W S$ 
for the $(N_f-K)$-th branch.}
\bea \label{equ-W-4}
0 & = & \log \left(\Sh^{-N_c} \right) + K \log
\left(\frac{\zh_f- \sqrt{\zh_f^2- {4\Sh }}  }{ 2}\right)
+(N_f-K) \log \left(
\frac{ \zh_f+ \sqrt{\zh_f^2- {4 \Sh }}  }{ 2}\right),
\eea
where we have defined dimensionless quantities 
$\Sh=\frac{S}{ m_A
\Lambda^2}= \frac{\W \mu}{ 4 \Lambda^2}$ 
and $\zh_f=\frac{z_f}{ \Lambda}$.
Note that using these massless quantities the cut is from along the
interval $[-2\sqrt{\Sh}, 2\sqrt{\Sh}]$.

Using (\ref{equ-W-1}) or (\ref{equ-W-4}) it is easy to show that
\bea 
W_{\text{eff,on-shell}} & = & \left(N_c-
\frac{N_f}{ 2}\right) S+ \frac{ m_A z_f^2}{ 4}
\left[ N_f +(2K-N_f) \sqrt{ 1-\frac{4 S}{ m_A z_f^2}} \right].
\nonu
\eea
Also by expanding $T(z)$ in the present case, just as we did in
\eqref{exact-T(z)}, we can read off
\bean
\braket{ \frac{m_A }{ 2} \Tr\, \Phi^2} 
 & = &
 \left(N_c-\frac{N_f}{ 2}\right) S + \frac{ m_A z_f^2}{ 4}
\left[ N_f+(2K-N_f) \sqrt{1-\frac{4 S}{ m_A z_f^2}} \right],
\eean
which gives us the relation
\bean
\braket{ \frac{m_A}{ 2} \Tr\, \Phi^2  }
={W_{\rm eff,on-shell}}
\eean
as we promised in (\ref{promise}).

Equation (\ref{equ-W-1}) is hard to solve. But if we want just to
discuss whether the cut closes up when we bring $z_f\to 0$, we can set
$K=0$ \footnote{If $K\neq 0$, we will have 
$U(N_c)\to U(K)\times
U(N_c-K)$ and the problem reduces to of 
$U(N_c-K)$ with $(N_f-2K)$
flavors in the 0-th branch.}, for which \eqref{equ-W-1} reduces to
\bea \label{eom_quad}
 z_f= \omega^r_{N_f}\Sh^{\frac{N_c}{ N_f}} +
 \omega^{-r}_{N_f} \Sh^{\frac{N_f-N_c}{ N_f}}.
\eea
Here, $\omega_{N_f}$ is the $N_f$-th root of unity,
$\omega_{N_f}=e^{2\pi i/N_f}$, and $r=0,1,\dots,(N_f-1)$ 
corresponds to
different branches of solutions.  It is also amusing to note that
above solution has the Seiberg duality \cite{Seiberg:1994pq} where
electric theory with $(N_c,N_f)$ is mapped to 
a magnetic theory with $(N_f-N_c,N_f)$.

With these preparations, we can start to discuss the on-shell process of
passing poles through the cut from the second sheet.

\subsection{Passing poles through a branch cut}

\indent

\begin{figure}[h]
\begin{center}
  \epsfxsize=7cm \epsfbox{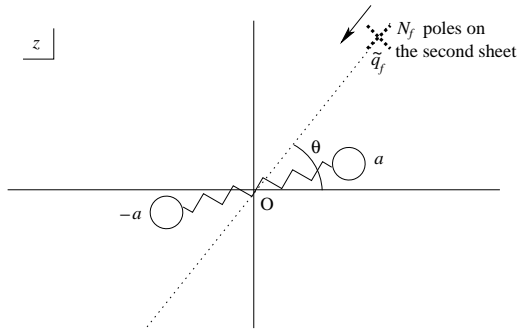}
\end{center}
 \vspace{-.5cm}
\caption{\sl A process in which $N_f$ poles at $\qt_f$ on the second
sheet far away from the cut approach the branch cut on the double
sheeted $z$-plane, along a line which goes through the origin and makes
an angle $\theta$ with the real $z$ axis.  The ``$\times$'' with dotted
lines denotes the poles on the second sheet, moving in the direction of
the arrow.  The two branch points $\pm a$ are connected by the branch
cut, which is denoted by a zigzag.  } \label{fig:poles_coming_in}
\end{figure}

Consider moving $N_f$ poles on top of each other at infinity on the
second sheet toward the cut along a line passing through the origin and
making an angle of $\theta$ with the real axis.  Namely,
take\footnote{Throughout this subsection, we will use the dimensionless
quantities ${\zh}_f$, $\Sh$, etc.\ and omit the hats on them to avoid
clutter, unless otherwise mentioned.}
\begin{align}
 z(\widetilde q_f)=z(q_f)=  p e^{i\theta}, \qquad p,\theta\in\mathbb{R}, 
 \label{move_pole_z(q)}
\end{align}
and change $p$ from $p=\infty$ to $p=-\infty$ (see Fig.\
\ref{fig:poles_coming_in}).  This equation \eqref{move_pole_z(q)} needs
some explanation.  Remember that $z_f=z(q_f)=z(\qt_f)$ denotes the
projection from the Riemann surface $\Sigma$ (Eq.\ \eqref{matrix-curve})
to the $z$-plane. For each point $z$ on the $z$-plane, there are two
corresponding points: $q$ on the first sheet and $\qt$ on the second
sheet.  Although we are starting with poles at $\qt_f$ on the second
sheet, we do not know in advance if the poles will pass through the cut
and end up on the first sheet, or it will remain on the second sheet.
Therefore we cannot specify which sheet the poles are on, and that is
why we used $z(q_f)$, $z(\qt_f)$ in \eqref{move_pole_z(q)}, instead of
$q_f$ or $\qt_f$.

Below, we study the solution to the equation of motion \eqref{eom_quad},
changing $p\in \mathbb{R}$ from $p=\infty$ to $p=-\infty$.  By
redefining $z_f$, $S$ by $z_f\to z_f e^{2\pi i r/(N_f-2N_c)}$, $S\to S
e^{4\pi i r/(N_f-2N_c)}$, we can bring \eqref{eom_quad} to the following
form:
\begin{align}
 z_f = p e^{i\theta}= S^{t}+S^{1-t},
\label{eom_quad(2)}
\end{align}
where
\begin{align}
 \frac{N_c}{ N_f}\equiv t.
\nonu
\end{align}
Henceforth we will use \eqref{eom_quad(2)}.  
Because $z_f$ as well as $S$ is complex, the position of the branch
points (namely, the ends of the cut), $\pm a$, where
\begin{align}
 a\equiv \sqrt{4S}
\nonu
\end{align}
is also complex, which means that in general the cut makes some finite
angle with the real axis, as shown in Fig.\ \ref{fig:poles_coming_in}.

\subsubsection*{\bf{$\bullet $ $\pmb{N_f=N_c}$}}

\indent

As the simplest example, let us first consider the $N_f=N_c$ (i.e.,
$t=1$) case.  We will see that the poles barely pass through the cut
but get soon bounced back to the second sheet.


The equation of motion \eqref{eom_quad(2)} is, in this case,
\begin{align}
 z_f=p e^{i \theta} = S + 1.
\label{eom_quad_Nf=N}
\end{align}
Therefore, as we change $p$, the position of the branch points changes
according to
\begin{align}
 a=\sqrt{4S}
 = 2\sqrt{z_f - 1} = 2\sqrt{pe^{i\theta} - 1}.
\label{pos_brch_pt}
\end{align}

Let us look closely at the process, step by step.  The point is that
transition between the first and the second sheet can happen only when
the cut becomes parallel to the incident direction of the poles, or when
the poles pass through the origin.
\begin{itemize}
 \item[(1)] $p\simeq +\infty$, on the second sheet:\\ 
            In this case, we can approximate the
            right hand side of \eqref{pos_brch_pt} as
            \begin{align*}
             a&
             =2p^{\half}e^{\frac{i\theta}{ 2}}(1-p^{-1}e^{-i\theta})^{\half}
             \simeq 2p^{\half}e^{\frac{i\theta}{ 2}} e^{-\half p^{-1}e^{-i\theta}}
             =2p^{\half}e^{-\half p^{-1}\cos{\theta}}\;
             e^{i(\frac{\theta}{ 2}+\half p^{-1}\sin\theta)}.
            \end{align*} 
            Therefore, when the poles are far away, the angle between
            the cut and the real axis is approximately $\frac{\theta}{
            2}>0$ (we assume $0<\theta<\frac{\pi}{ 2}$).  Furthermore,
            as the poles approach ($p$ becomes smaller), the cut
            shrinks (because of $p^{\half}$) and rotates
            counterclockwise (because of $e^{\frac{i}{
            2}p^{-1}\sin\theta}$).  This corresponds to Fig.\
            \ref{fig:cfg_cut_poles}a.
            
 \item[(2)] Because the cut is rotating counterclockwise, as the poles
            approach, the cut will eventually become parallel to the
            incident direction, at some point.  This
            happens when
            \begin{align*}
             a^2&=4(pe^{i\theta}-1)=4\left[(p\cos\theta-1)+ip\sin\theta\right] 
             \propto e^{2i\theta} = \cos 2\theta+i\sin 2\theta.
            \end{align*}
            By simple algebra, one
            obtains
            \begin{align}
             p=2\cos\theta, \qquad a=2e^{i\theta}.\label{horzntl_pt}
            \end{align}
            Note that this is the only solution; the cut becomes
            parallel to the incident direction only once. Because
            $0<p<|a|=2$ (we are assuming $0 < \theta < \pi/2$), by the
            time the cut becomes parallel to the incident direction, the
            poles have come inside of the interval $[-2e^{i\theta},
            2e^{i\theta}]$, along which the cut extends when it is
            parallel to the incident direction.
	    This implies that the poles cross \footnote{As mentioned in
	    footnote \ref{phys_mean_cut}, there is no real physical
	    meaning to the position of the cut itself.} the cut at this
	    point, and enter into the first sheet.
            Fig.\ \ref{fig:cfg_cut_poles}b shows the situation when this
            transition is about to happen.
            In Fig.\ \ref{fig:cfg_cut_poles}c,  the poles are
            just crossing the cut.
            Fig.\ \ref{fig:cfg_cut_poles}d corresponds to the situation just
            after the transition happened; the poles have passed the cut
            and are now proceeding on the first sheet.
            
            Here we implicitly assumed that the cut is still rotating
            counterclockwise with a finite angular velocity, but this
            can be shown by expanding $a$ around \eqref{horzntl_pt} as
            $p=2\cos\theta+\Delta p$.  A short computation shows
            \begin{align*}
             a&\simeq 2  e^{\half \Delta p \, \cos\theta}
             e^{i\theta-\frac{i}{ 2} \Delta p \, \sin\theta}
            \end{align*}
            which implies that the cut shrinks and rotates
            counterclockwise if we move the poles to the left ($\Delta p$
            decreases).
            
\begin{figure}[ht]
\begin{center}
  \epsfxsize=16cm \epsfbox{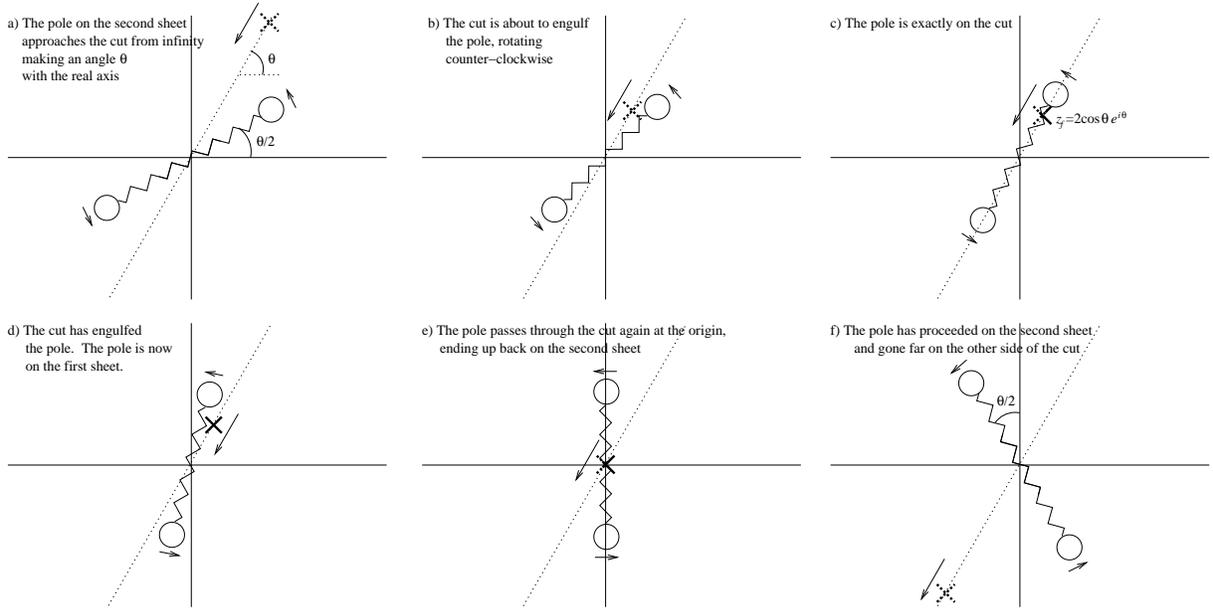}
\end{center}
\begin{center}
\caption{\sl Six configurations of the branch cut and the poles.  The
poles are depicted by ``$\times$'' and moving along a line at an angle
$\theta$ with the real axis, as the arrow on it indicates.  The
``$\times$'' in solid (dotted) lines denotes poles on the first (second)
sheet.  The branch cut is rotating counterclockwise (as the arrows on
its sides indicate), changing its length.  }
\label{fig:cfg_cut_poles}
\end{center}
\end{figure}

 \item[(3)] $p\simeq 0$:\\ If the poles proceed on the real line
            further, it eventually reaches the origin $p=0$.
            By expanding \eqref{pos_brch_pt} around $p=0$, one obtains
            \begin{align}
             a&= 2(e^{\pi i}+pe^{i\theta})^{\half}
             \simeq 2e^{-\half p \cos\theta} e^{i(
             \frac{\pi}{ 2}-\half p \sin\theta)}.
             \label{Nf=Nc,zf=0}
            \end{align}
            Therefore the cut has a finite size ($|a|=2$) at $p=0$ and
            along the imaginary axis, still rotating counterclockwise, but now
            expanding. Because the cut goes through the origin, the poles
            pass through the cut again and comes back onto the second
            sheet (Fig.\ \ref{fig:cfg_cut_poles}e).

 \item [(4)] $p\simeq -\infty$, on the second sheet:\\ 
            If the poles have gone far past the cut so that $p<0$, $|p|\gg
            1$, we can approximate \eqref{pos_brch_pt}, as before, as 
            \begin{align*}
             a
             &= 2|p|^{\half}e^{\frac{i}{2}(\theta+\pi)}
             (1-p^{-1}e^{-i\theta})^{\half}
             \simeq 
             2|p|^{\half}e^{\half |p|^{-1}\cos{\theta}}\;
             e^{i(\frac{\pi}{ 2}+\frac{\theta}{2}-
             \half |p|^{-1}\sin\theta)}.
            \end{align*}
            Therefore, as the poles go away, the cut expands and
            rotates counterclockwise.  The angle between the cut and the
            real axis asymptotes to $(\frac{\pi}{2}+
            \frac{\theta}{2})$
            (Fig.\ \ref{fig:cfg_cut_poles}f).

\end{itemize}

In the above we assumed that $0<\theta<\frac{\pi}{2}$.  If
$\frac{\pi}{2}<\theta<\pi$, 
the only difference is that the order of steps
(2) and (3) are exchanged.  If $\theta<0$, the cut rotates clockwise
instead of counterclockwise.

When is the cut shortest in this whole process?
From \eqref{pos_brch_pt}, one easily obtains 
\begin{align}
 & |a|=2[(p-\cos\theta)^2+\sin^2\theta]^{1/4}\ge 2|\sin\theta|^{1/2}.
 \label{cut_length}
\end{align}
Therefore, when the poles are at $z_f=pe^{i\theta}=\cos\theta\,
e^{i\theta}$, which is between the steps (2) and (3) above, the cut
becomes shortest.  In particular, in the limit $\theta\to 0$ or
$\theta\to \pm\pi$, the cut completely closes up instantaneously.  These
correspond to configurations with either a horizontal cut with poles
colliding sideways, or a vertical cut with poles colliding from right
above or from right below.  Actually the existence of the $S=0$ solution
is easy to see in \eqref{eom_quad_Nf=N}: it is just $z_f=1,S=0$.

Summary: for $N_f=N_c$, when one moves poles on the second sheet from
infinity along a line toward a cut, poles pass through the cut onto
the first sheet and move away from the cut by a short distance. 
Then 
poles are bounced back to the second sheet again.  Therefore, one can
never move poles far away from the cut on the first sheet.  During
the process, in certain situations, the cut completely closes up.

\subsubsection*{\bf{$\bullet $ $\pmb{N_f\neq N_c}$}}

\indent

Now let us consider a more general case with $N_f\neq N_c$.  We again
consider a situation where poles on the second sheet approach a cut.
This time we will be brief and sketchy, because a detailed analysis such
as the one we did for the $N_f=N_c$ case would be rather lengthy due to
the existence of multiple branches, and would not be very illuminating.

First, let us ask how we can see whether poles are on the first sheet or
on the second sheet, from the behavior of $S$ versus $p$.  
Because this
is not apparent in the equation of motion of the form
\eqref{eom_quad(2)}, let us go back to
\begin{align}
 0=\frac{\partial W_{\rm eff}}{ \partial  S}
 \propto
 \log S^{-N_c}+{N_f}\log 
 \frac{z_f\mp \sqrt{z_f^2-4S}}{ 2}
 ~~~~\Longrightarrow~~~~
 S^{t}
 =\frac{z_f\mp \sqrt{z_f^2-4S}}{ 2}.
 \label{eom_quad_pre}
\end{align}
which led to the equation \eqref{eom_quad(2)}.  Here the ``$-$''
(``$+$'') sign corresponds to $q_f$ on the first (second) sheet. For
$|z_f|^2\gg |4S|$, the square root can be approximated as
$\sqrt{z_f^2-4S}=z_f(1-4S/z_f^2)^{1/2}\simeq z_f(1-2S/z_f^2)$ (our sign
convention was discussed above \eqref{quadratic-exact}).  Therefore
\eqref{eom_quad_pre} is, on the first sheet,
\begin{align}
 S^{t} \simeq  \frac{z_f-z_f(1-2S/z_f^2)}{ 2}
 = \frac{S}{ z_f}
 ~~~~~~\Longrightarrow~~~~~~
 z_f\simeq S^{1-t}, 
 \label{S-q_sht1}
\end{align}
while on the second sheet
\begin{align}
 S^t \simeq \frac{z_f+z_f(1-2S/z_f^2)}{ 2}
 \simeq  {z_f}
 ~~~~~~\Longrightarrow~~~~~~
 z_f\simeq S^t.
 \label{S-q_sht2}
\end{align}

Now let us solve \eqref{eom_quad(2)} for $|z_f|\gg 1$.  By carefully
comparing the magnitude of the two terms in \eqref{eom_quad(2)}, one
obtains
\begin{align}
 \begin{array}{ccccl@{}c}
 N_f<N_c&(1<t) &\rightarrow&
  \left\{
  \begin{array}{l}
   z_f\simeq S^t \\[1ex]
   z_f\simeq S^{1-t}
  \end{array}
  \right.
  &
  \begin{array}{@{}l}
   \rightarrow~~ |S|\simeq |p|^{\frac{1}{ t}},    \\[1ex]
    \rightarrow~~ |S|\simeq |p|^{-\frac{1}{ t-1}},
  \end{array}
  &
  \begin{array}{c}
    |S|\gg 1    \\[1ex]
    |S|\ll 1
  \end{array}
  \\[4ex]
 N_c<N_f<2N_c&(\thalf<t<1) &\rightarrow&
 z_f\simeq S^t 
 & \rightarrow~~ |S|\simeq |p|^{\frac{1}{ t}}, &|S|\gg 1 \\[2ex]
 2N_c<N_f&(0<t<\thalf) &\rightarrow& z_f\simeq S^{1-t} 
 & \rightarrow~~ |S|\simeq |p|^{\frac{1}{ 1-t}}, & |S|\gg 1
\end{array}
 \label{large-q_behavior}
\end{align}
It is easy to show that $|z_f|^2\gg |4S|$ in all cases. So by using
\eqref{S-q_sht1}, \eqref{S-q_sht2}, we conclude that the first and the
third lines in \eqref{large-q_behavior} correspond to poles on the
second sheet, while the second and the last lines correspond to poles
on the first sheet.  This implies that, only for $N_f<N_c$, poles on
the second sheet can pass through the cut 
all the way and go infinitely
far away on the first sheet from a cut, 
as we will see explicitly in the
examples below.  
For $N_c<N_f<2N_c$, if one tries to pass poles through a cut, then
either poles will be bounced back to the second sheet, or the cut
closes up before the poles reach it.  
For $N_f>2N_c$, there is no solution corresponding to poles moving
toward a cut from infinity on the second sheet.  This should be related
to the fact that in this case glueball $S$ is not a good IR field.
Therefore, this one cut model is not applicable for $N_f>2N_c$.

To argue that these statements are true, rather than doing an analysis
similar to the one we did for the $N_f=N_c$ case, we will present some
explicit solutions for some specific values of $t=\frac{N_c}{ N_f}$ and
$\theta$, and argue general features.  Before looking at explicit
solutions, note that there are multiple solutions to Eq.\
\eqref{eom_quad(2)} which can be written as
\begin{align}
 z_f=(S^{\frac{1}{ N_f}})^{N_c}+(S^{\frac{1}{ N_f}})^{N_f-N_c}.
\label{eom_quad_ito_T}
\end{align}
From the degree of this equation, one sees that the  number of
the solutions to \eqref{eom_quad_ito_T} is:
\begin{align}
\begin{array}{ccl}
 N_f\le N_c     &\Longrightarrow & \text{$2N_c-N_f$ solutions,}\\
 N_c\le N_f\le 2N_c&\Longrightarrow & \text{$N_c$ solutions,}\\
 2N_c\le N_f    &\Longrightarrow & \text{$N_f-N_c$ solutions.}
\end{array}
\nonu
\end{align}
%
Therefore, if we solve the equation of motion \eqref{eom_quad(2)}, in
general we expect multiple branches of solutions.  We do not have to
consider the $N_f$ branches of the root $S^{\frac{1}{ N_f}}$, because it
is taken care of by the phase rotation we did above \eqref{eom_quad(2)}.

\begin{figure}[ht]
 \vspace{.5cm}
 \begin{center}
  \begin{tabular}{ccc}
   \epsfxsize=4cm \epsfbox{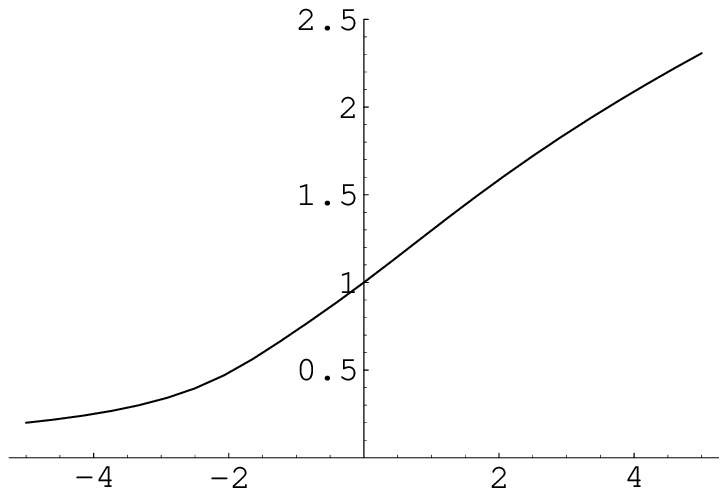}&
   \epsfxsize=4cm \epsfbox{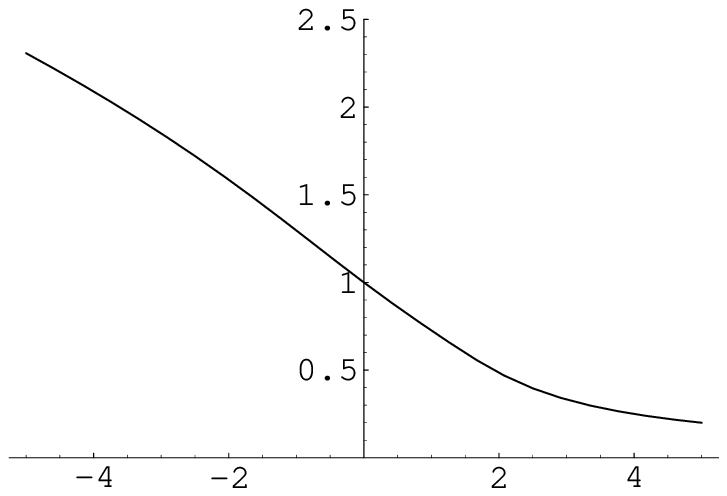}&
   \epsfxsize=4cm \epsfbox{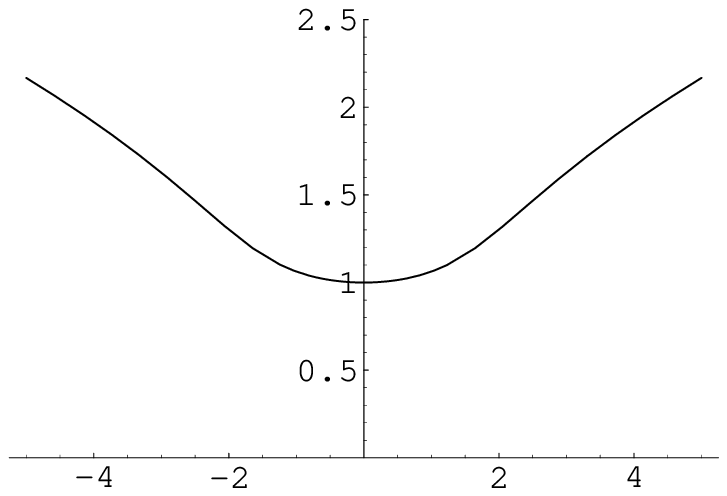}\\
   branch i) & branch ii) &branch iii) \\
  \end{tabular}
  \caption{\sl The graph of $|S|$ versus $p$ for ${N_f}=\half N_c$
  ($t=2$), $\theta=\pi/6$.
  The vertical axis is $|S|$ and the horizontal axis is $p$.
  Although we are showing just the $\theta=\pi/6$ case, there are
  similar looking three branches for any value of $\theta$, which change
  into one another when $\theta$ is changed by $2\pi/3$.
  }
  \label{fig:S-q_rf=1/2}
 \end{center}
\end{figure}

Now let us look at  explicit solutions for $U(2)$ example.
For ${N_f}=\half N_c$ ($t=2$), the solution to the equation of motion
\eqref{eom_quad(2)} is
\begin{align}
 S&=
 -\frac{1}{ 3}
 \left(
\frac{27+\sqrt{729-108z_f^3}}{ 2}\right)^{\!\!\!1/3}\!\!\!
 - \left(\frac{27+\sqrt{729-108z_f^3}}{ 2}\right)^{\!\!\!-1/3},
\nonu
\end{align}
where three branches of the cubic root are implied.
In Fig.\ \ref{fig:S-q_rf=1/2} we plotted $|S|$ versus $p$ for these
branches, for a randomly chosen value of the angle of incidence,
$\theta=\pi/6$.  Even if one changes $\theta$, there are always three
branches whose general shapes are similar to the ones in Fig.\
\ref{fig:S-q_rf=1/2}.  These three branches changes into one another
when $\theta$ is changed by $2\pi/3$.
One can easily see which
branch corresponds to what kind of processes, 
by the fact that on the
first sheet $|S|\ll 1$ as $|p|\to \infty$, while on the second sheet
$|S|\gg 1$ as $|p|\to \infty$.
The three branches correspond to the process in which: 
i)  poles go from the second sheet to the first sheet through the
cut, without any obstruction,
ii)  poles go from the first sheet to the second sheet (this is not
the process we are interested in), and
iii)  poles coming from the second sheet get reflected back to the
second sheet.
Note that the cut has never closed 
in all cases, because $|S|$ is always
nonvanishing.

%
%
%
%

\begin{figure}[ht]
 \vspace{.5cm}
 \begin{center}
  \begin{tabular}{cc}
   \epsfxsize=4cm \epsfbox{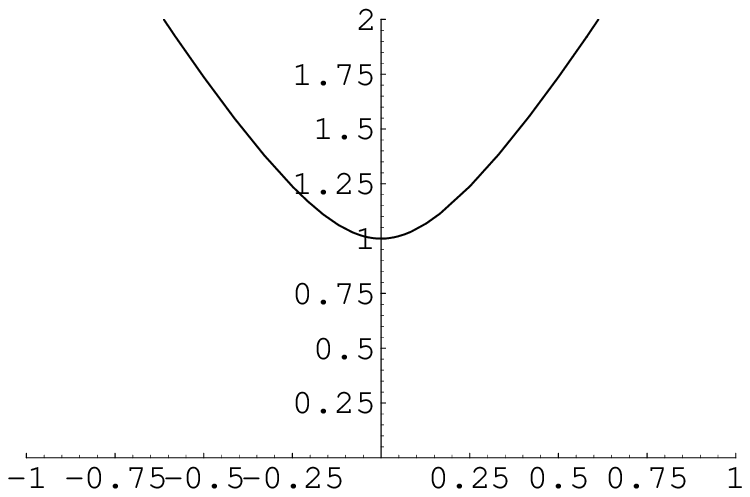}&
   \epsfxsize=4cm \epsfbox{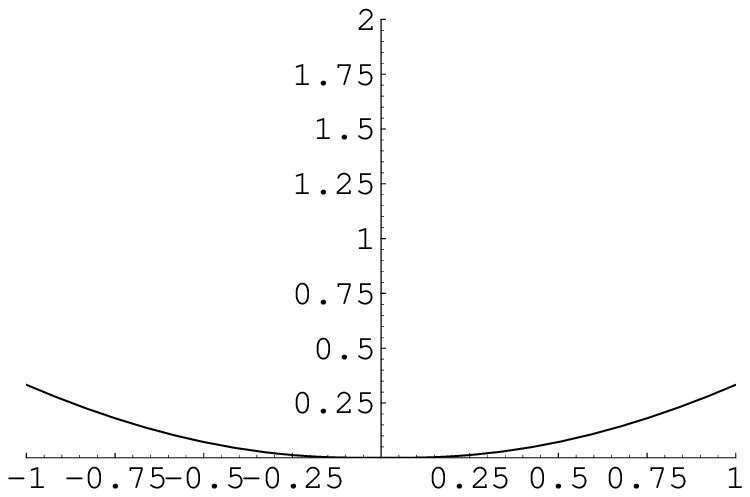}\\
   branch i) & branch ii)
  \end{tabular}
 \caption{\sl The graph of $|S|$ versus $p$ for ${N_f}=\frac{3}{ 2}N_c$
  ($t=\frac{2}{ 3}$) or $N_f=3N_c$ ($t=\frac{1}{ 3}$), for
  $\theta=\pi/2$.  The two values of $t$ give the same graph.  The
  vertical axis is $|S|$ and the horizontal axis is $p$.  Although we
  are showing just the $\theta=\pi/2$ case, there are similar looking
  two branches for any value of $\theta$, which change into each other
  when $\theta$ is changed by $\pi$,
  } \label{fig:S-q_rf=3/2}
 \end{center}
\end{figure}

Similarly, for ${N_f}=\frac{3}{ 2}N_c$ ($t=\frac{2}{ 3}$), 
the solution to the equation of motion
\eqref{eom_quad(2)} is
\begin{align}
 S&=\frac{1}{ 2}\left[-3z_f-1\pm(z_f+1)\sqrt{4z_f+1}\right].
\nonu
\end{align}
This time there are two branches, which change into each other when the
$\theta$ is changed by $\pi$.  We plotted $|S|$ versus $p$ for
$\theta=\pi/2$ in Fig.\ \ref{fig:S-q_rf=3/2}.  It shows two
possibilities: i) poles coming from the second sheet get reflected
back to the second sheet, for which $|S|\neq 0$ as $p\to 0$, ii) the cut
closes up before poles passes through it, for which $|S|\to 0$ as $p\to
0$.

%
%
%
%


The $N_f=3N_c$ ($t=\frac{1}{ 3}$) case is also described by the same
Fig.\ \ref{fig:S-q_rf=3/2}.  However, as we discussed below
\eqref{large-q_behavior}, it does not correspond to a process of poles
approaching the cut from infinity on the second sheet; it corresponds to
poles on the first sheet and we cannot give any physical interpretation
to it.

\bigskip
These demonstrate the following general features: 
\begin{itemize}
 \item For $N_f< N_c$, one can move poles at infinity on the second
       sheet through a cut all the way to infinity on the first sheet
       without obstruction, if one chooses the incident angle
       appropriately.  If the angle is not chosen appropriately, the
       poles will be bounced back to the second sheet.

 \item For $N_c\le N_f<2N_c$, one cannot move poles at infinity on the second
       sheet through a cut all the way to infinity on first sheet.
       If one tries to, either i) the cut rotates and sends the poles
       back to the second sheet, or ii) the cut closes up before the
       poles reach it.

 \item For $2N_c<N_f$, the one cut model does not apply directly.
\end{itemize}

%

$N_f=2N_c$ is an exceptional case, 
for which the equation of
motion \eqref{eom_quad(2)} becomes
\begin{align}
 z_f=2 S^{1/2}.
\label{sol_Nf=2Nc}
\end{align}
Therefore $|S|\to 0$ as $z_f\to 0$, and the cut always closes before the
poles reach it.  

\subsubsection*{Subtlety in $S=0$ solutions}

If $p=0$, or equivalently if $z_f=0$, there is a subtle, but important
point we overlooked in the above arguments.  For $z_f=0$, the
superpotential \eqref{superpot_quad} becomes
\begin{align}
 W_{\rm eff}
 &=
 S\left[\left( N_c-\frac{N_f}{ 2}\right)
 +\log\left(
\frac{(-1)^{N_f/2}m_A^{N_c-N_f/2}
\Lambda^{2N_c-N_f}}{ S^{N_c-N_f/2}}\right)
 \right]
 \notag\\
 &=
 S\left[\left( N_c-\frac{N_f}{ 2}\right)+
 \log\left(
\frac{(-1)^{-N_f/2}m_A^{N_c-N_f/2}
\Lambda_0^{2N_c-N_f}}{ S^{N_c-N_f/2}}\right)
 \right]
 +2\pi i \tau_0 S\notag\\
 &=
 \left( N_c-\frac{N_f}{ 2}\right)S
\left[1+ \log \left(\frac{\widetilde\Lambda_0^3}{ S} \right)
 \right]
 +2\pi i \tau_0 S
 \label{Weff_zf=0}
\end{align}
with $\widetilde\Lambda_0^{3(N_c-N_f/2)}\equiv
(-1)^{-N_f/2}m_A^{N_c-N_f/2}\Lambda_0^{2N_c-N_f}$.  Only from here to
\eqref{q=0_exclusion}, $S$ means the dimensionful quantity
($S=m_A\Lambda^2 \Sh$; see below \eqref{equ-W-4}).  In addition, in the
second line of \eqref{Weff_zf=0}, we rewrite the renormalized scale
$\Lambda$ in terms of the bare scale $\Lambda_0$ and the bare coupling
$\tau_0$ using the relation \eqref{scale}.  In our case,
$B_L=(-1)^{N_f}$.
The equation of motion derived from \eqref{Weff_zf=0}
is
\begin{align}
 \left( N_c-\frac{N_f}{ 2}\right)
\log \left(\frac{\widetilde\Lambda_0^3}{ S} \right) 
+ 2\pi i \tau_0
 =0
\nonu
\end{align}
and the solution is
\begin{align}
 S=
 \begin{cases}
 \widetilde\Lambda_0^3 \,e^{\frac{2\pi i \tau_0}{ N_c-N_f/2}}   
& N_f\neq 2N_c~, \\
 \text{no solution}  & N_f=2N_c~.
 \end{cases}
 \label{q=0_exclusion}
\end{align}

For $N_f<N_c$, \eqref{q=0_exclusion} is consistent with the fact that
the $|S|$ versus $p$ graphs in Fig.\ \ref{fig:S-q_rf=1/2} all go through
the point $(p,|S|)=(0,1)$ (now $S$ means the dimensionless quantity).
Also for $N_f=N_c$, \eqref{q=0_exclusion} is consistent with the result
\eqref{Nf=Nc,zf=0} ($|a|=2$, so $|S|=1$).  On the other hand, for
$N_f>N_c$, \eqref{q=0_exclusion} implies that we should exclude the
origin $(p,|S|)=(0,0)$ from the $|S|$-$p$ graphs in Fig.\
\ref{fig:S-q_rf=3/2}, which is the only $S=0$ solution (this includes
the $N_f=2N_c$ case \eqref{sol_Nf=2Nc}).  \footnote{One may think that
if one uses the first line of \eqref{Weff_zf=0}, then for $N_f=2N_c$,
$W_{\rm eff}=S\log[(-1)^{N_f/2}]$ and there are solutions for some
$N_f$.  However, the glueball superpotential that string theory predicts
\cite{CIV,Ookouchi} is the third line of \eqref{Weff_zf=0} which is in
terms of the bare quantities $\Lambda_0$ and $\tau_0$.  If $N_f=2N_c$,
then the log term vanishes and one cannot define a new scale $\Lambda$
as we did in \eqref{scale} to absorb the linear term $2\pi i \tau_0 S$.}

Therefore, the above analysis seems to indicate that, for $N_f>N_c$, the
$S=0$ solution at $p=0$, or equivalently $z_f=0$ is an exceptional case
and should be excluded. On the other hand, as can be checked easily,
gauge theory analysis based on the factorization method shows that there
is an $S=0$ solution in the baryonic branch. Thus we face the problem of
whether the baryonic $S=0$ branch for $N_f>N_c$ can be described in
matrix model, as alluded to in the previous discussions.

Note that, there is also an $S=0$ solution for $N_c=N_f$ in certain
situations, as discussed below \eqref{cut_length}.  For this solution,
which is in the non-baryonic branch, there is no subtlety in the
equation of motion such as \eqref{q=0_exclusion}, and it appears to be a
real on-shell solution.  This will be discussed further below.

\subsection{Generalization of IKRSV for the one cut model}
\label{subsec:general'n_IKRSV}

\indent

In the above and in subsection \ref{subsec:prospects_strong_cpl}, we
argued that for $N_f\geq N_c$ the $S=0$ solutions are real, on-shell
solutions based on the factorization analysis.  More accurately, there
are two cases with $S=0$: the one in the maximal non-baryonic branch
with $N_f\geq 2N_c$ and the other one in the baryonic branch with
$N_c\le N_f < 2N_c$. The case of non-baryonic branch cannot be discussed
in the one cut model, which is applicable only to $N_f<2N_c$.  On the
other hand, the baryonic one did show up in the previous subsection, but
we just saw above that those solutions should be excluded by the matrix
model analysis.  What is happening?
Is it impossible to describe the baryonic branch in matrix model?

Recall that the glueball field $S$ has to do with the strongly coupled
dynamics of $U(N_c)$ theory.  That $S=0$ in those solutions means that
there is no strongly coupled dynamics any more, namely the $U(N_c)$
group has broken down completely.  The only mechanism for that to happen
is by condensation of a massless charged particle which makes the $U(1)$
photon of the $U(N_c)$ group massive.
Therefore, in order to make $S=0$ a solution, we should incorporate such
an extra massless degree of freedom, which is clearly missing in the
description of the system in terms only of the glueball $S$.
This extra degree of freedom should exist even in the $N_f=N_c$ case
where $S=0$ really is an on-shell solution as discussed below
\eqref{cut_length}; we just could not directly see the degree of freedom
in this case.

The analysis of \cite{IKRSV} hints on what this extra massless degree of
freedom should be in the matrix model / string theory context.  Note
that, the superpotential \eqref{Weff_zf=0} is of exactly the same form
as Eq.\ (4.5) of \cite{IKRSV}, if we interpret $N_c-N_f/2\equiv \Nh$ as
the amount of the net RR 3-form fluxes.
In \cite{IKRSV} it was argued that, if the net RR flux $\Nh$ vanishes,
one should take into account an extra degree of freedom corresponding to
D3-branes wrapping the blown up $S^3$ in the Calabi--Yau geometry
\cite{Strominger:1995cz}, and condensation of this extra degree of
freedom indeed makes $S=0$ a solution to the equation of motion.  The
form of the superpotential \eqref{Weff_zf=0} strongly suggests that the
same mechanism is at work for $N_f=2N_c$ in the $r=N_c$ non-baryonic
branch; condensation of the D3-brane makes $S=0$ a solution.
Furthermore, as discussed in \cite{IKRSV}, for $N_f>2N_c$ the glueball
$S$ is not a good variable and should be set to zero.  A concise way of
summarizing this conclusion is: if the generalized dual Coxeter number
$h=N_c-N_f/2$ is zero or negative, we should set $S$ to zero in the
$r=N_c$ non-baryonic branch.

However, this is not the whole story, as we have discussed in subsection
\ref{subsec:prospects_strong_cpl}.  As we saw above, we need some extra
physics also for $N_c\le N_f<2N_c$ in order to explain the matrix model
result in the baryonic branch.
We argue below that this extra degree of freedom at least in the $N_c <
N_f<2N_c$ case should also be the D3-brane wrapping $S^3$ which shrinks
to zero when the glueball goes to zero: $S\to 0$.

The original argument of \cite{IKRSV} is not directly applicable for
$N_f<2N_c$ because there are nonzero RR fluxes penetrating such a
D3-brane ($\Nh\neq 0$).  These RR fluxes induce fundamental string
charge on the D3-brane.  Because the D3-brane is compact, there is no
place for the flux to end on (note that this flux is not the RR one but
the one associated with the fundamental string charge).  Hence it should
emanate some number of fundamental strings.  If there are no flavors,
there is no place for such fundamental strings to end on, so they should
extend to infinity.  This fact led to the conclusion of \cite{IKRSV}
that the D3-brane wrapping $S^3$ is infinitely massive and not relevant
unless $\Nh=0$.

However, in our situation, there are places for the fundamental strings
to end on --- noncompact D5-branes which give rise to flavors
\cite{CIV,Ookouchi}.  In particular, precisely in the $z_f=0$ case,
where we have $S=0$ solutions for $N_c<N_f<2N_c$, the D3-brane wrapping
$S^3$ intersects the noncompact D5-branes in the $S\to 0$ limit, hence
the 3-5 strings stretching between them are massless.  Therefore the
D3-brane with these fundamental strings on it is massless and should be
included in the low energy description.
It is well known \cite{Witten:1998xy} that such a D-brane with
fundamental strings ending on it can be interpreted as baryons in gauge
theory.\footnote{That the D3-brane wrapping $S^3$ cannot exist for
$N_f<N_c$ can probably be explained along the same line as
\cite{Witten:1998xy}, by showing that those 3-5 strings are fermionic.
Also, note that the gauge group here is $U(N_c)$, not $SU(N_c)$ as 
in \cite{Witten:1998xy}, hence the ``baryon'' is charged under the $U(1)$.
  }
Condensation of this baryon degree of freedom should make $S=0$ a
solution, making the photon massive and breaking the $U(N_c)$ down to
$U(0)$.  The precise form of the superpotential for this extra degree of
freedom must be more complicated than the one proposed in \cite{IKRSV}
for the case without flavors.
%

%
%

All these analyses tell us the following prescription: 
\begin{align}
 \label{prescription}
 \qquad
 \begin{minipage}{5.5in}
  \baselineskip=18pt
  \em
  Using the floating mass condition that all $N_f$ poles are on top of
  one branch point\/%
  \addtocounter{footnote}{1}%
  \protect\footnotemark[\arabic{footnote}] 
  on the Riemann surface, we will have 
  an $S=0$ solution for $N_f\geq 2N_c$. For $N_c< N_f<2N_c$ there are two
  solutions: one with $S=0$ in the baryonic branch and one with
  $S\neq 0$ in the non-baryonic branch. In multi-cut cases, this 
  applies to each cut by replacing $N_c,$ $S$ with the corresponding
  $N_{c,i}$, $S_i$ for the cut.
 \end{minipage}
\end{align}
\footnotetext[\arabic{footnote}]{This condition will not work for the
$N_f=N_{c,i}$ case.}

In the next section we will discuss the condition we have used
in above prescription. Also by explicit examples, we
will demonstrate that when the gauge theory has a solution with closed
cuts ($S_i=0$), one can reproduce its superpotential in matrix model by
setting the corresponding glueballs $S_i$ to zero by hand.

  %
  %


\section{Two cut model---cubic tree level superpotential} 
\setcounter{equation}{0}
\label{sec:2cut}

\indent

Now, let us move on to $U(N_c)$ theory with cubic tree level
superpotential, where we have two cuts.  We will demonstrate that 
for each closed cut we can set $S=0$ by hand to 
reproduce the correct gauge theory superpotential using matrix model.

Specifically, we take the tree level superpotential to be
\begin{equation}
 \begin{split}
 W_{\rm tree}&=\Tr[W(\Phi)]-\sum_{I=1}^{N_f} \widetilde{Q}_I (\Phi-z_f) Q^I,\\
 W(z)&=\frac{g}{ 3}z^3+\frac{m}{ 2}z^2,\qquad
 W'(z)=gz\left(z+\frac{m}{ g}\right)\equiv g(z-a_1)(z-a_2). 
\end{split}
\label{hnhn8Apr04}
\end{equation}
Here we wrote down $W(z)$ in terms of $g_2=g$, $g_1=m$ for definiteness,
but mostly we will work with the last expression in terms of $g$,
$a_{1,2}$.
The general breaking pattern in the pseudo-confining phase is $U(N_c)\to
U(N_{c,1})\times U(N_{c,2})$, $N_{c,1}+N_{c,2}=N_c$, $N_{c,i}>0$.  In the quantum theory, the
critical points at $a_1$ and $a_2$ blow up into cuts along the intervals
$[a_1^-,a_1^+]$ and $[a_2^-,a_2^+]$, respectively.  Namely, we end up
with the matrix model curve \eqref{csw2-2.18}, which in this case is
\begin{align}
 y_m^2= W'(z)^2+f_1(z)= g^2(z-a_1^-)(z-a_1^+)(z-a_2^-)(z-a_2^+).
\label{hpni8Apr04}
\end{align}
We will call the cuts along $[a_1^-,a_1^+]$ and $[a_2^-,a_2^+]$
respectively the ``first cut'' and the ``second cut'' henceforth.  One
important difference from the quadratic case is that, we can study a
process where $N_f\ge 2N_{c,i}$ flavor poles are near the $i$-th cut in
the cubic case. 

%

As we have mentioned, our concern is whether the cut is closed or not.
Also from the experiences in the factorization it can be seen that for
$N_f>N_{c,i}$, when closed cut is produced, the closed cut and the poles are
on top of each other \footnote{We do not discuss the $N_f=N_{c,i}$ case
where closed cut and poles are not at the same point.  However because the
$S=0$ solution in this case is an on-shell solution, we can reproduce
the gauge theory result in matrix model without setting $S=0$ by hand.}.
With all these considerations we take the following condition to
constrain the position of the poles:\footnote{We could choose
$z_f=a_1^+$ or $z_f=a_2^\pm$ instead of \eqref{zf=a1-}, but the result
should be all the same, so we take \eqref{zf=a1-} without loss of
generality.}
\begin{align}
 z_f=a_1^-.
\label{zf=a1-}
\end{align}
If there are $S_1=0$ solutions in which the closed cut and the poles are
on top of each other, then all such solutions can be found by solving
the factorization problem under the constraint \eqref{zf=a1-}, since for
such solutions $z=a_1^-=a_1^+$ obviously.  One could impose a further
condition $S_1=0$, or equivalently $a_1^-=a_1^+$ if one wants just
closed cut solutions, but we would like to know that there also are
solutions with $S_1\neq 0$ for $N_f<2N_{c,1}$, so we do not do that.

To summarize, what we are going to do below is: first we  explicitly solve
the factorization problem under the constraint \eqref{zf=a1-}, and
confirm that the $S_1=0$ solution exists  when $N_f>N_{c,i}$. 
 Then, we reproduce the gauge theory
superpotential in matrix model by setting $S_1=0$ by hand.

\bigskip Before plunging into that, we must discuss one aspect of the
constraint \eqref{zf=a1-} and the $r$-branches, in order to understand
the result of the factorization method.  If one solves the factorization
equation for a given flavor mass $m_f=-z_f$ (without imposing the
constraint \eqref{zf=a1-}), then in general one will find multiple
$r$-branches labeled by an integer $K$ with range $0\le K\le {\rm
min}(N_c,[\frac{N_f}{2}])$ (see Eq.\ \eqref{range_r-vac}).  This is
related to the fact that the factorization method cannot distinguish
between the poles on the first sheet and the ones on the second sheet.
The $r$-branch labeled by $K$ corresponds to distributing $N_f-K$ poles
on the second sheet and $K$ poles on the first sheet.  We are not
interested in such configurations; we want to put $N_f$ poles at the
same point on the same sheet.  However, as we discuss now, we actually
do not have to worry about the $r$-branches under the constraint
\eqref{zf=a1-}.

\begin{figure}[h]
\begin{center}
  \epsfxsize=7cm \epsfbox{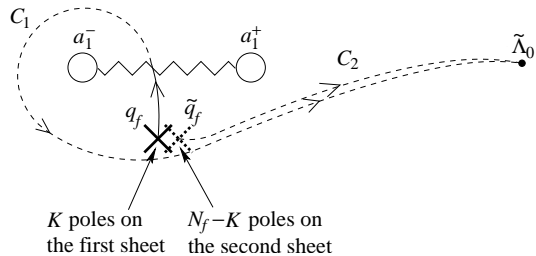}
\end{center}
\vspace{-.5cm}
\caption{\sl Contours $C_1$ and $C_2$ defining $\Pi_f^{(1)}$ and
 $\Pi_f^{(2)}$, respectively.  The part of a contour on the first sheet
 is drawn in a solid line, while the part on the second sheet is drawn
 in a dashed line.  The $N_f-K$ poles on the second sheet and the $K$
 poles on the first sheet are actually on top of each other (more
 precisely, their projections to the $z$-plane are.)}
 \label{fig:contours}
\end{figure}

The $r$-branches with different $K$ are different vacua in
general. However, under the constraint \eqref{zf=a1-} these $r$-branches
become all identical because at the branch point $z=a_i^\pm$ there is no
distinction between the first and second sheets.  This can be easily
seen in the matrix model approach.  From the equation (\ref{Wexact-y}),
the effective glueball superpotential for the two cut model with $N_f-K$
poles at $\qt_f$ on the second sheet and $K$ poles at $q_f$ on the first
sheet is
\begin{align*}
 W_{\rm eff}& = -\frac{1}{ 2} ( N_{c,1} \Pi_1+ N_{c,2} \Pi_2)
 -\frac{1}{ 2} (N_f-K) \Pi_f^{(2)}
 -\frac{1}{ 2} K \Pi_{f} ^{(1)}
 +\frac{1}{ 2}(2N_c-N_f) W(\Lambda_0) \\ 
 &\quad 
 +\frac{1}{ 2} N_f W(q)  -\pi i(2N_c-N_f)S+ 2\pi i \tau_0 S
 +2\pi i  b_1 S_1,
\end{align*}
where the periods are defined by
\begin{align*}
 S_i  &=  \frac{1}{ 2\pi i }\int_{A_{i}} R(z) dz,\qquad
 \Pi_i  =  2\int_{a_{i}^-}^{\Lambda_0} y(z) dz,\\
 \Pi_f^{(2)}  &=  \int_{\qt_f}^{\W\Lambda_0} y(z) dz, \qquad
 \Pi_{f}^{(1)} 
 =  \int_{q_f}^{\W \Lambda_0} y(z) dz
 =\left[\int_{q_f}^{\W q_f} + \int_{\W q_f}^{\W\Lambda_0}\right] y(z) dz
 \equiv \Delta \Pi_f +\Pi_f^{(2)} 
\end{align*}
with $i=1,2$.  The periods $\Pi_f^{(1)}$, $\Pi_f^{(2)}$ are associated
with the poles on the first sheet and the ones on the second sheet,
respectively.  The contour $C_2$ for $\Pi_f^{(2)}$ is totally on the
second sheet, while the contour $C_1$ for $\Pi_f^{(1)}$ is from $q_f$
on the first sheet, through a cut, to $\W\Lambda_0$ on the second sheet.
These contours are shown in Fig.\ \ref{fig:contours}.
This $r$-branch with $K$ poles on the first sheet can be reached by
first starting from the pseudo-confining phase with all $N_f$ poles at
$\qt_f$ ($K=0$) and then moving $K$ poles through the cut to $q_f$.  The
path along which the poles are moved in this process is the difference
in the contours, $C_1-C_2\equiv \Delta C$ \footnote{There is ambiguity
in taking $\Delta C$; for example we can take $\Delta C$ to go around
$a_1^+$ in Fig.\ \ref{fig:contours}.  However, the difference in
$\int_{\Delta C} y(z)dz$ for such different choices of $\Delta C$ is
$2\pi i n S_1$, $n\in \mathbb{Z}$, which can be absorbed in redefinition
of the theta angle and is immaterial.}.

\begin{figure}[h]
\begin{center}
  \epsfxsize=7cm \epsfbox{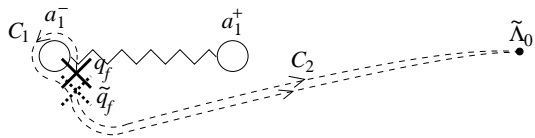}
\end{center}
 \vspace{-.5cm}
\caption{\sl Under the constraint \eqref{zf=a1-}, contours $C_1$ and
 $C_2$ become degenerate: $C_1=C_2$.}  \label{fig:contours_deg}
\end{figure}

When we impose the constraint \eqref{zf=a1-}, then the difference
$\Delta C$ vanishes (Fig.\ \ref{fig:contours_deg}).  Therefore there is
no distinction between $C_1$, $C_2$ and hence $\Pi_f^{(1)}=\Pi_f^{(2)}$
for any $K$.  In other words, all $K$-th branches collapse\footnote{In
fact this collapse was observed in \cite{afo1,afo2} for $SO(N_c)$ and
$USp(2N_c)$ gauge groups with massless flavors. We have seen that there
are only two branches, i.e., Special branch and Chebyshev branch, which
correspond to the baryonic branch and the non-baryonic branch in
$U(N_c)$ case.}  to the same branch under the constraint \eqref{zf=a1-}.

\bigskip Now, let us explicitly solve the factorization problem under
the constraint \eqref{zf=a1-}, and check that the $S=0$ solutions exist
as advertised before.  In solving the factorization problem, we do not
have to worry about the $r$-branches because there is no distinction
among them under the constraint \eqref{zf=a1-}.  Then, we compute the
exact superpotential using the data from the factorization and reproduce
it in matrix model by setting $S=0$ by hand when $S=0$ on the gauge
theory side.

For simplicity and definiteness, we consider the case of $U(3)$ gauge
group henceforth.  We consider $N_f<2N_c$ flavors, namely $1\le N_f\le 5$,
because $N_f\ge 2N_c$ cases are not asymptotically free and cannot be
treated in the framework of the Seiberg--Witten theory.

\subsection{Gauge theory computation of superpotential}
\label{QuantumMassless}

\indent

In this subsection, we solve the factorization equation under the
constraint \eqref{zf=a1-} and compute the exact superpotential, for the
system (\ref{hnhn8Apr04}) with $U(3)$ gauge group and with various
breaking patterns.  

\subsubsection{Setup}

\indent

The factorization equation for $U(3)$ theory with $N_f$ flavors with mass
$m_f=-z_f$ is given by\footnote{From the result of Appendix
(\ref{Nocorrection}), the matrix model curve with flavors does not
change even for $N_f > N_c$,
contrary to [7].}  \cite{bfhn}
\bea
 \widetilde{P}_3(\W z)^2-4\La^{6-N_f}(\W z-z_f)^{N_f}
&= & \W H_{1}(\W z)^2\left[W^{\prime}(\W z)^2+f_{1}(\W z) \right]\nonumber\\
&= &\W H_{1}(\W z)^2 (\W z-a_1^-)(\W z-a_1^+)(\W z-a_2^-)(\W z-a_2^+) ,\label{cvn11Apr04}
\eea
where we set $g=1$ for simplicity and $W^{\prime}(\W z)$ is given by
(\ref{hnhn8Apr04}).  The breaking pattern is assumed to be $U(3)\to
U(N_{c,1})\times U(N_{c,2})$ with $N_{c,i}>0$.
Here we used new notations to clarify the shift of the coordinate below.
For quantities after the shift, we use letters without tildes. Enforcing
the constraint \eqref{zf=a1-}
and shifting
$\zt$ as $\W z=z+a_1^-$, we can rewrite this relation as follows:
\bea
P_3(z)^2-4\La^{6-N_f}z^{N_f}&= &
H_{1}(z)^2\left[ \W W^{\prime}(z)^2+ \W f_1(z)\right]=
H_{1}(z)^2\left[z(z-\W a_1^+)(z-\W a_2^-)(z-\W a_2^+) \right]\nonumber\\
&\equiv & H_{1}(z)^2 \left[z(z^3+Bz^2+Cz+D) \right].
\label{factorization} 
\eea Because of the shift, the polynomials $P_3(z)$ and $H_1(z)$ are
different in form from $\W P_3(\W z)$ and $\W H_1(\W z)$ in
(\ref{cvn11Apr04}).  We parametrize the polynomials $P_3(z)$ and
$H_1(z)$ as
\begin{eqnarray}
P_3(z)=z^3+az^2+bz+c,\qquad  H_1(z)=z-A. 
\label{p3h1}
\end{eqnarray}
The parameters $B, C$ can be written in terms of the parameters in
\bea
\W W^{\prime}(z)=  W^{\prime}(\W z)=\W z(\W z+m_A)=
(z+a_1^-)(z+a_1^-+m_A)\equiv (z-a_1) (z-a_2) 
\nonu
\eea
by comparing the coefficients in
(\ref{hpni8Apr04}):
\bea
 a_1&= &\frac{-B \mp {\sqrt{3\,B^2 - 8\,C}}}{4},\ \  
a_2=\frac{-B \pm {\sqrt{3\,B^2 - 8\,C}}}{4}, \nonumber \\
\Delta^2 &\equiv & (\W a_2-\W a_1)^2=(a_2-a_1)^2= \frac{{{3\,B^2 - 8\,C}}}{4}.
 \label{constraint}
\eea
The ambiguity in signs in front of the square roots can be fixed by
assuming $a_1<a_2$. Finally we undo the shift by noting that 
\bea
W(\W z) & = & \frac{1}{ 3} \W z^3+\frac{m_A}{ 2} \W z^2
 =  
 \frac{z^3}{ 3}-(a_1+a_2)\frac{z^2}{ 2}+a_1 a_2 z
 + \frac{1}{ 6}(a_1^3-3a_1^2 a_2). \label{field-form}
\eea

With all this setup, we can compute the superpotential as follows.
First we factorize the curve according to (\ref{factorization}).  Then
we find the Casimirs $U_1, U_2, U_3$ from $P_3(z)$ \footnote{From the
coefficients of $P_3(z)$, namely $a$, $b$ and $c$, one can compute the
Casimirs $U_k=\frac{1}{ k}\braket{\Tr\, \Phi^k}$ using the quantum
modified Newton relation Appendix (\ref{PN2}) as explained in Appendix
\ref{comp_spot-gt}.}  and solve for $a_1, a_2$ using the last equation
of (\ref{constraint}). Finally we put all these quantities into
(\ref{field-form}) to get the effective action as
\bea
 W_{\rm low}
 =\braket{\Tr\, W(\Phi) }
 &= & 
 U_3 - (a_1+a_2)U_2 + a_1 a_2  U_1 
 + \frac{N_c}{ 6}(a_1^3-3a_1^2 a_2).
 \label{W_low_U(3)}
\eea
Here, for $a_{1}$, $a_2$, one can use the first two equations of
\eqref{constraint}.  We also want to know whether the first cut (the one
along the interval $[a_1^-,a_1^+]$) is closed or not; we expect that the
cut closes if we try to bring too many poles near the cut.  As is
obvious from \eqref{factorization}, this can be seen from the value of
$D$. If $D=0$, the cut is closed, while if $D\neq 0$, the cut is open.

\subsubsection{The result of factorization problem}

\begin{table}
\begin{center}
 \begin{tabular}{||c||c||c|c|c||c||} 
  \hline
  $N_f$  & 
  \begin{minipage}{1.25in}\begin{center}breaking pattern\\$\widehat{U(N_{c,1})}\times U(N_{c,2})$\end{center}\end{minipage}  & 
      $N_f\le N_{c,1}$ & $N_{c,1}<N_f<2N_{c,1}$ & $2N_{c,1}\le N_f$&
  \begin{minipage}{.75in}\begin{center}the first\\cut is\end{center}\end{minipage} \\  \hline\hline
 $1$ &  $\widehat{U(2)}\times U(1)$ & $\bigcirc$ & -          & -         & open    \\ \cline{2-6}
     &  $\widehat{U(1)}\times U(2)$ & $\bigcirc$ & -          & -         & open    \\ \hline
 $2$ &  $\widehat{U(1)}\times U(2)$ & -          & -          & $\bigcirc$& closed  \\ \cline{2-6}
     &  $\widehat{U(2)}\times U(1)$ & $\bigcirc$ & -          & -         & open    \\ \hline
 $3$ &  $\widehat{U(1)}\times U(2)$ & -          & -          & $\bigcirc$& closed  \\ \cline{2-6}
     &  $\widehat{U(2)}\times U(1)$ & -          & $\bigcirc$ & -         & closed  \\ \cline{2-6}
     &  $\widehat{U(2)}\times U(1)$ & -          & $\bigcirc$ & -         & open    \\ \hline
 $4$ &  $\widehat{U(1)}\times U(2)$ & -          & -          & $\bigcirc$& closed  \\ \cline{2-6}
     &  $\widehat{U(2)}\times U(1)$ & -          & -          & $\bigcirc$& closed  \\ \hline
 $5$ &  $\widehat{U(1)}\times U(2)$ & -          & -          & $\bigcirc$& closed  \\ \cline{2-6}
     &  $\widehat{U(2)}\times U(1)$ & -          & -          & $\bigcirc$& closed  \\ \hline
\end{tabular} 
\begin{center}
\caption{\sl The result of factorization of curves for $U(3)$ with up to
$N_f=5$ flavors.  ``$\,\bigcirc$'' denotes which inequality $N_f$ and $N_{c,1}$ satisfy.
}
\end{center}
\end{center}
\end{table}

\indent

We explicitly solved the factorization problem for $U(3)$ gauge theories
with $N_f=1,2,\dots, 5$ and summarized the result in Table 1.  Let us
explain about the table.  
$\widehat{U(N_{c,1})}\times U(N_{c,2})$ denotes the breaking pattern of the
$U(3)$ gauge group.  The hat on the first factor means that the pole is
at one of the branch points of the first cut (Eq.\ \eqref{zf=a1-}) which
is associated with the first factor $U(N_{c,1})$. 
Of course this choice is arbitrary and
we may as well choose $U(N_{c,2})$, ending up with the same result.
Finally, whether the cut is closed or not depends on whether $D=0$ or
not, as explained below \eqref{W_low_U(3)}.

%
%
%

Now let us look carefully at Table 1, comparing it with the prescription
\eqref{prescription} based on the analysis of the one cut model.

First of all, for $N_f\ge 2N_{c,1}$, the first cut is always closed.  We
will see below that in these cases with a closed cut the superpotential
can be reproduced by setting $S_1=0$ by hand in the corresponding matrix
model, confirming the prescription \eqref{prescription} for
$N_f>2N_{c,1}$.


Secondly, the lines for $N_f=3$ and $\widehat{U(2)}\times U(1)$
correspond to the $N_{c,1}<N_f<2N_{c,1}$ part of the prescription
\eqref{prescription}.  There indeed are both an open cut solution and a
closed cut solution.
We will see below that the superpotential of the closed cut solution can
be reproduced by setting $S_1=0$ by hand in the corresponding matrix
model.  On the other hand, the superpotential of the open cut solution
can be reproduced by not setting $S_1=0$, namely by treating $S_1$ a
dynamical variable and extremizing $W_{\rm eff}$ with respect to it.
In fact these two solutions are baryonic branch for a closed cut and
non-baryonic branch for an open cut.

Finally, for $N_f\le N_{c,1}$, the cut is always open, which is also
consistent with the prescription \eqref{prescription}.  In this case,
the superpotential of the open cut solution can be reproduced by
extremizing $W_{\rm eff}$ with respect to it, as we will see below.
For $N_f=N_{c,1}$ there should be an $S_1=0$ ($a_1^-=a_1^+$) solution
for some $z_f$ (corresponding to $U(2)$ theory with $N_f=2$ in the $r=0$
branch) in the quadratic case, but under the constraint \eqref{zf=a1-}
we cannot obtain that solution.

%
%
%


\bigskip
Below we present  resulting exact superpotentials, for all possible
breaking patterns.  For simplicity, we do not take care of phase factor
of $\La$ which gives rise to the whole number of vacua.  For details of
the calculation, see Appendix \ref{comp_spot-gt}.

\subsubsection*{Results}

\indent

Definitions: $ W_{\rm cl}= -\frac{1}{3}$ for $\widehat{U(1)}\times
U(2)$ and $ W_{\rm cl}= -\frac{1}{6}$ for $\widehat{U(2)}\times
U(1)$.  For simplicity we set $g=1$ and $\Delta=a_2-a_1=-m/g=1$.

\begin{itemize}
 \item ${\bf \widehat{U(1)}\times U(2)}$ {\bf with } ${\bf N_f=1}$
       \begin{eqnarray}
        W_{\rm low}
         = W_{\rm cl} - 2T - \frac{5T^2}{2} + 
         \frac{115 T^3}{12} - \frac{245T^4}{4} + 
         \frac{30501 T^5}{64} - \frac{12349 T^6}{3}+\cdots,
         \quad 
         T\equiv  \La^{\frac{5}{2}}.
          \nonu
       \end{eqnarray}

 \item  ${\bf \widehat{U(2)}\times U(1)}$ {\bf with } ${\bf N_f=1}$ 
       \begin{eqnarray}
        W_{\rm low}
         = W_{\rm cl} - \frac{5T^2}{2} + \frac{5T^3}{3} - 
         \frac{11T^4}{3} + 11T^5 - \frac{235T^6}{6}+\cdots,
         \quad T\equiv \Lambda ^{\frac{5}{3}}.  \nonu
       \end{eqnarray}

 \item ${\bf \widehat{U(1)}\times U(2)}$ {\bf with } ${\bf N_f=2}$ 
       \begin{eqnarray}
        W_{\rm low}
         = W_{\rm cl} + 2T^2 - 6T^4 - \frac{32T^6}{3} - 
         40T^8 - 192T^{10} - \frac{3136T^{12}}{3}+\cdots,
         \quad T \equiv  \La. \nonu
       \end{eqnarray}

 \item ${\bf \widehat{U(2)}\times U(1)}$ {\bf with } ${\bf N_f=2}$
       \begin{eqnarray}
        W_{\rm low}
         = W_{\rm cl} - 2T^4 - \frac{16T^6}{3} - 24T^8 - 
         128T^{10} - \frac{2240T^{12}}{3}+\cdots,\quad 
         T\equiv  \Lambda.  \nonu
       \end{eqnarray}

 \item  ${\bf \widehat{U(1)}\times U(2)}$ {\bf with } ${\bf N_f=3}$ 
       \begin{eqnarray}
        W_{\rm low}
         = W_{\rm cl} + 2T - \frac{19T^2}{2} + 
         \frac{51T^3}{4} + \frac{157T^4}{4} + 
         \frac{5619T^5}{64} + \frac{33T^6}{2}+\cdots,\quad 
         T\equiv  \Lambda^{\frac{3}{2}}.  \nonu
       \end{eqnarray}

 \item  ${\bf \widehat{U(2)}\times U(1)}$ {\bf with } ${\bf N_f=3}${\bf : two solutions}
       \begin{eqnarray}
        W_{\rm low,baryonic}
         &=&W_{\rm cl} + T - \frac{5T^2}{2} - 33T^3 - 
         543T^4 - 10019T^5 - \frac{396591T^6}{2}+\cdots,
         T\equiv  \La^3.  \nonu
         \\
        W_{\rm low}
         &=& W_{\rm cl} + {\Lambda }^3.\nonu
       \end{eqnarray}

 \item  ${\bf \widehat{U(1)}\times U(2)}$ {\bf with } ${\bf N_f=4}$ 
       \begin{eqnarray}
        W_{\rm low}
         &=& W_{\rm cl} + 2T - 13T^2 + \frac{176T^3}{3} - 
         138T^4 + 792T^6 \nonu \\ &&- 9288T^8 + 
         137376T^{10} 
        - 2286144T^{12}+\cdots,\quad
         T\equiv  \Lambda.  \nonu
       \end{eqnarray}

 \item ${\bf \widehat{U(2)}\times U(1)}$ {\bf with } ${\bf N_f=4}$ 
       \begin{eqnarray}
        W_{\rm low}
         &=& W_{\rm cl} + T - 6T^2 - \frac{40T^3}{3} - 
         56T^4 - 288T^5 - \frac{4928T^6}{3}
        -9984T^7 - 63360T^8 \nonu \\&& - \frac{1244672T^9}{3} - 
         2782208T^{10} - 19009536T^{11}\cdots,\quad
         T\equiv \Lambda^2 . \nonu
       \end{eqnarray}

 \item  ${\bf \widehat{U(1)}\times U(2)}$ {\bf with } ${\bf N_f=5}$ 
       \begin{eqnarray}
        W_{\rm low}
         & = & W_{\rm cl} - 2T - \frac{33T^2}{2} - 
         \frac{1525T^3}{12}- \frac{3387T^4}{4}   - 
         \frac{314955T^5}{64} \nonu \\
         && - \frac{74767T^6}{3}+\cdots,
         \quad T\equiv \Lambda ^{\frac{1}{2}}. \nonu
       \end{eqnarray}
 \item ${\bf \widehat{U(2)}\times U(1)}$ {\bf with } ${\bf N_f=5}$ 
\begin{eqnarray}
W_{\rm low}
= W_{\rm cl}+ T - \frac{19T^2}{2} + 
  \frac{154T^3}{3} - 132T^4 + 828T^6+\cdots,\quad 
  T\equiv \Lambda.  \nonu
\end{eqnarray}
\end{itemize}

\subsection{Matrix model computation of superpotential}

\indent

In this subsection we compute the superpotential of the system
\eqref{hnhn8Apr04} in the framework of \cite{CSW-II}.  If all the $N_f$
flavors have the same mass $m_f=-z_f$, the effective glueball
superpotential $W_{\rm eff}(S_j)$ for the pseudo-confining phase with
breaking pattern $U(N_c)\to \prod_{i=1}^n U(N_{c,i})$, $\sum_{i=1}^n
N_{c,i}=N_c$ is, from \eqref{Wexact-y},
\begin{align}
 W_{\rm eff}(S_j)
 &=
 -\frac{1}{ 2}\sum_{i=1}^n N_{c,i} \Pi_i - \frac{N_f}{ 2} \Pi_f
 +\left(N_c-\frac{N_f}{ 2}\right)W(\Lambda_0) 
 +\frac{N_f}{ 2}W(z_f)
 \cr
 &\qquad
 -2\pi i\left(N_c-\frac{N_f}{ 2}\right)S
 +2\pi i\tau_0 S
 +2\pi i\sum_{i=1}^{n-1} b_i S_i,\label{jziq11Apr04}
\end{align}
where the periods associated with adjoint and fundamentals are defined
by
\begin{gather*}
 \Pi_i(S_j)\equiv 2 \int_{a_i^-}^{\Lambda_0} y(z) dz,\qquad
 \Pi_f(S_j)\equiv \int_{\W z_f}^{\widetilde\Lambda_0} y(z) dz 
  = -  \int_{z_f}^{\Lambda_0} y(z) dz,\\
 y(z)=\sqrt{W'(z)^2+f_1(z)}~.
\end{gather*}

For cubic tree level superpotential \eqref{hnhn8Apr04},
the periods $\Pi_{1,2}(S_j)$ were computed by explicitly evaluating the
period integrals by power expansion in \cite{CIV}, as
\begin{align}
 \frac{\Pi_1}{ 2g\Delta^3}=\,&
 \frac{1}{ g\Delta^3}[W(\Lambda_0)-W(a_1)]
 +s_1\left[1+\log\left(\frac{\lambda_0^2}{s_1}\right)\right]
 +2s_2\log \lambda_0\cr
 &+(
 -2s_1^2
 +10s_1s_2
 -5s_2^2
 )
 +\left( 
   -\frac{32}{ 3} s_1^3
   +91 s_1^2 s_2
   -118 s_1 s_2^2
   + \frac{91}{ 3}s_2^3
 \right)
 \cr
 &+\left( 
   -\frac{280}{3}s_1^4
   +\frac{3484}{3}s_1^3s_2
   - 2636s_1^2 s_2^2
   + \frac{5272}{3}s_1 s_2^3
   - \frac{871}{3}s_2^4 
 \right)
 +\cdots,
 \cr
 \frac{\Pi_2}{ 2g\Delta^3}=\,&
 \frac{1}{ g\Delta^3}[W(\Lambda_0)-W(a_2)]
 +s_2\left[1+\log\left(\frac{\lambda_0^2}{-s_2}\right)\right]
 +2s_1\log \lambda_0\notag\\
 &
 +(2s_2^2-10s_1s_2+5s_1^2)
 +\left(
 -\frac{32}{ 3}s_2^3
 +91s_1s_2^2
 -118s_1^2s_2
 +\frac{91}{ 3}s_1^3
 \right)
 \notag\\
 &
 +\left(
 \frac{280}{3}s_2^4
 -\frac{3484}{3}s_1s_2^3
 +2636s_1^2s_2^2
 -\frac{5272}{3}s_1^3s_2
 +\frac{871}{3}s_1^4
 \right)
 +\cdots,\label{jwhq8Apr04}
\end{align}
where $\Delta\equiv a_2-a_1$, $s_i\equiv S_i/g\Delta^3$, and
$\lambda_0\equiv \Lambda_0/\Delta$.


Under the constraint \eqref{zf=a1-}, the contours defining $\Pi_1$ and
$\Pi_f$ coincide, so
\begin{align}
 \frac{1}{ 2}\Pi_1=
-\Pi_f=\int_{z_f=a_1^-}^{\Lambda_0}y(z)dz.\label{mqes8Apr04}
\end{align} 
Using this, we can rewrite \eqref{jziq11Apr04} as
\begin{align}
 W_{\rm eff}(S_1,S_2)
 &=
 \Bigl[N_{c,1} W(a_1)+N_{c,2} W(a_2)\Bigr]
  -\frac{N_f}{ 2} \Bigl[W(a_1)-W(z_f)\Bigr]
 \cr 
 &\qquad
 -{\Nt_{c,1}} \Bigl[\half\Pi_1-W(\Lambda_0)+W(a_1)\Bigr]
 -N_{c,2} \Bigl[\half\Pi_2-W(\Lambda_0)+W(a_2)\Bigr]
 \cr
 &\qquad
 -2\pi i(\Nt_{c,1}+N_{c,2})S+2\pi i\tau_0 S  +2\pi ib_1 S_1.
 \label{juho8Apr04}
\end{align}
Here we rearranged the terms taking into account the fact that the
periods take the form $\half\Pi_i=W(\Lambda_0)-W(a_i)+\text{(quantum
correction of order $\CO(S_i)$)}$, and also the fact that we are
considering $z_f=a_1^-\simeq a_1$ (thus the second term).  The first line
corresponds to the classical contribution, while the second and third
lines correspond to quantum correction.  Furthermore, we defined
$\Nt_{c,1}\equiv N_{c,1}-N_f/2$.

We would like to extremize this $W_{\rm eff}$ \eqref{juho8Apr04} with
respect to $S_{1,2}$, and compute the low energy superpotential that can
be compared with the $W_{\rm low}$ obtained in the previous subsection
using gauge theory methods.
In doing that, one should be careful to the fact that one should treat
the mass $z_f$ as an external parameter which is independent of
$S_{1,2}$ although we are imposing the constraint \eqref{zf=a1-},
$z_f=a_1^-=a_1^{-}(S_1,S_2)$.  Where is the $z_f$ dependence in
\eqref{juho8Apr04}?  Firstly, $z_f$ appears explicitly in the second
term in \eqref{juho8Apr04}.  Therefore, when we differentiate $W_{\rm
eff}$ with respect to $S_{1,2}$, we should exclude this term.  Secondly,
there is a more implicit dependence on $z_f$ in $\Pi_f =
-\int_{z_f}^{\Lambda_0} y(z)dz$, which we replaced with $-\Pi_1/2$ 
using
\eqref{mqes8Apr04}.  If we forget to treat $z_f$ as independent of
$S_i$, then we get an apparently unwanted, extra contribution as
$\frac{\partial \Pi_f }{ \partial S_i} = -\int_{z_f}^{\Lambda_0}
\frac{\partial y(z) }{\partial S_i} dz -\frac{\partial z_f}{ \partial
S_i}\cdot y(z)|_{z=z_f} $.  However, this last term actually does not
make difference because
\begin{align*}
y(z)|_{z=z_f}=g\left.\sqrt{(z-z_f)(z-a_1^+)(z-a_2^-)(z-a_2^+)}\right|_{z=z_f}=0.
\end{align*}
Therefore what one should do is: i) plug the expression
\eqref{jwhq8Apr04} into \eqref{juho8Apr04}, ii) solve the equation of
motion for $S_{1,2}$ using \eqref{juho8Apr04} without the second term,
and then iii) substitute back the value of $S_{1,2}$ into
\eqref{juho8Apr04}, now with the second term included.

Solving the equation of motion can be done by first writing the
Veneziano--Yankielowicz term (log and linear terms) as
\begin{align*}
 & -{\Nt_{c,1}} \Bigl[\half\Pi_1-W(\Lambda_0)+W(a_1)\Bigr]
 -N_{c,2} \Bigl[\half\Pi_2-W(\Lambda_0)+W(a_2)\Bigr]
 \\
 &\qquad\qquad\qquad\qquad\qquad\qquad
 -2\pi i(\Nt_{c,1}+N_{c,2})S+2\pi i\tau_0 S  +2\pi ib_1 S_1
  \\
 &\qquad\qquad
 = 
 g\Delta^3\biggl\{ \Nt_{c,1} s_1 [ 1-\log(s_1/\lambda_1^3)] +  N_{c,2} s_2 [ 1-\log(s_2/\lambda_2^3)] 
 + \CO(s_i^2)
 \biggr\},
\end{align*}
where
\bea
 \lambda_1^{3\Nt_{c,1}} 
 & =& 
\lambda_0^{2(\Nt_{c,1}+2N_{c,2})} e^{2\pi i (\Nt_{c,1}+N_{c,2})-2\pi i \tau_0-2\pi i b_1},\nonu \\
 \lambda_2^{3N_{c,2}} & = & 
(-1)^{N_{c,2}}\lambda_0^{2\Nt_{c,1}+2N_{c,2}}
 e^{2\pi i (\Nt_{c,1}+N_{c,2})-2\pi i \tau_0},
\nonu
\eea
and then solving the equation of motion perturbatively in
$\lambda_{1,2}$.
In this way, one can straightforwardly reproduce the results obtained in
the previous section in the case with the first cut {\em open\/}.  In the
case with the first cut {\em closed\/}, in order to reproduce the results
in the previous section, one should first set $S_1=0$ by hand, and then
extremize $W_{\rm eff}$ with respect to the remaining dynamical variable
$S_2$.
%

Following the procedure above, we checked explicitly that
extremizing $W_{\rm eff}(S_1,S_2)$ (open cut) or $W_{\rm
eff}(S_1\!=\!0,S_2)$ (closed cut) reproduces the $W_{\rm low}$ up the order
presented in the previous section, for all breaking patterns for $U(3)$
theory.

\bigskip 
In the above, we concentrated the explicit calculations of effective
superpotentials in $U(3)$ theory with cubic tree level superpotential.
These explicit examples are useful to see that the prescription
\eqref{prescription} really works; one can first determine using
factorization method when we should set $S_i=0$ by hand, and then
explicitly check that the superpotential obtained by gauge theory can be
reproduced by matrix model.

However, if one wants only to show the equality of the two effective
superpotentials on the gauge theory and matrix model sides, one can
actually prove it in general cases.  In Appendix \ref{CVmethod}, we
prove this equivalence for $U(N_c)$ gauge theory with an degree $(k+1)$
tree level superpotential where $k+1<N_c$.  There, we show the
following: if there are solutions to the factorization problem with some
cuts closed, then the superpotential $W_{\rm low}$ of the gauge theory
can be reproduced by extremizing the glueball superpotential $W_{\rm
eff}(S_i)$ on the matrix model side, after setting the corresponding
glueball fields $S_i$ to zero by hand.  Note that, on the matrix side we
do not know when we should set $S_i$ to zero {\it a priori\/}; we can
always set $S_i$ to zero in matrix model, but that does not necessarily
correspond to a physical solution on the gauge theory side that solves
the factorization constraint.

\section{Conclusion and some remarks}
\setcounter{equation}{0}
\label{sec:conclusion}


\indent

In this paper, taking $\CN=1$ $U(N_c)$ gauge theory with an adjoint and
flavors, we studied the on-shell process of passing $N_f$ flavor poles
on top of each other on the second sheet through a cut onto the first
sheet.  This corresponds to a continuous transition from the
pseudo-confining phase with $U(N_c)$ unbroken to the Higgs phase with
$U(N_c-N_f)$ unbroken (we are focusing on one cut).  We confirmed the
conjecture of \cite{CSW-II} that for $N_f<N_c$ the poles can go all the
way to infinity on the first sheet, while for $N_f\ge N_c$ there is
obstruction.  There are two types of obstructions: the first one is that
the cut rotates, catches  poles and sends them back to the first
sheet, while the second one is that the cut closes up before  poles
reach it.  The first obstruction occurs for $N_c\le N_f<2N_c$ whereas
the second one occurs for $N_c<N_f $.

If a cut closes up, the corresponding glueball $S$ vanishes, which means
that the $U(N_c)$ group is completely broken down.  This can happen only
by condensation of a charged massless degree of freedom, which is
missing in the matrix model description of the system.  With a massless
degree of freedom missing in the description, the $S=0$ solution should
be singular in matrix model in some sense.  Indeed, we found that the
$S=0$ solution of the gauge theory does not satisfy the equation of
motion in matrix model (with an exception of the $N_f=N_c$ case, where
the $S=0$ solution does satisfy the equation of motion).  How to cure
this defect of matrix model is simple --- the only thing the missing
massless degree of freedom does is to make $S=0$ a solution, so we just
set $S=0$ by hand in matrix model.  We gave a precise prescription
\eqref{prescription} when we should do this, i.e.,
{\em in the baryonic branch for $N_{c,i}\leq N_f< 2N_{c,i}$ and in the
$r=N_{c,i}$ non-baryonic branch\/}, and checked it with specific
examples.

The string theory origin of the massless degree of freedom can be
conjectured by generalizing the argument in \cite{IKRSV}.  We argued
that it should be the D3-brane wrapping the blown up $S^3$, along with
fundamental strings emanating from it and ending on the noncompact
D5-branes in the Calabi--Yau geometry.

Although we checked that the prescription works, the string theory
picture of the $S=0$ solution needs further refinement, which we leave
for future research.
For example, although we argued that some extra degree of freedom makes
$S=0$ a solution, we do not have the precise form of the superpotential
including that extra field.  It is desirable to derive it and show that
$S=0$ is indeed a solution, as was done in \cite{IKRSV} in the case
without flavors.
%
%
Furthermore, we saw that there is an on-shell $S=0$ solution for
$N_f=N_c$.  Although this solution solves the equation of motion in
matrix model, there should be a massless field behind the scene.  It is
interesting to look for the nature of this degree of freedom.  It cannot
be the D3-branes with fundamental strings emanating from it,
since for this solution the noncompact D5-branes are at finite distance
from the collapsed $S^3$ and the 3-5 strings are massive.
Finally, we found that the $S=0$ solution is in the baryonic branch.  It
would be interesting to ask if one can describe the baryonic branch in
the matrix model framework by adding some extra degrees of freedom.

\bigskip
In the following, we study some aspects of the theory, which we could
not discuss so far.  
We will discuss generalization to $SO(N_c)$ and $USp(2N_c)$ gauge groups
by computing the effective superpotentials with quadratic tree level
superpotential.

\subsection{$SO(N_c)$ theory with flavors}
\setcounter{equation}{0}

\indent

Here we consider the one cut model for $SO(N_c)$ gauge theory with
$N_f$ flavors.
The tree level superpotential of the theory 
is obtained from ${\cal N}=2$ SQCD  by adding the mass 
$m_A$ for the adjoint scalar $\Phi$
\bea
W_{\rm tree} = \frac{m_A}{2} \Tr\, \Phi^2 + Q^{f} \Phi Q^{f'}
J_{ff'} + 
Q^f \widetilde{m}_{ff'} Q^{f'}.
\label{tree}
\eea
where $f=1, 2, \cdots, 2N_f$ and the 
symplectic metric $J_{ff'}$ and 
mass matrix for quark $\widetilde{m}_{ff'}$
are given by 
\bea
J=\left(\begin{array}{cc} 0 & 1 \\ -1 & 0 \end{array} \right)
\otimes {\bf I}_{N_f\times N_f}, \qquad
\widetilde{m}=\left(\begin{array}{cc} 0 & 1 \\ 1 & 0 \end{array} 
\right)
\otimes \mbox{diag} (m_1, \cdots, m_{N_f})~.
\nonu
\eea 
For this simple case the matrix model curve is given by
\bea
y(z)^2= m_A^2 \left( z^2 - 4 \mu^2 \right).  
\label{soy}
\eea
This Riemann surface  is a double cover of the 
complex $z$-plane branched at the roots of $y_m^2$ (that is 
$z=\pm 2 \mu$).

The effective superpotential receives 
contributions from both the sphere and the disk amplitudes 
in the matrix model \cite{seiberg02}  
and the explicit form was given in \cite{CSW-II} for $U(N_c)$
gauge theory with flavors. 
Now we apply this procedure to our $SO(N_c)$ gauge theory
with flavors and it turns out the following expression  
\bea
W_{\rm eff} & = & -\frac{1}{2} \left( \sum_{i=-n}^{n} N_{c,i} -
2 \right) \int_{\widehat{B}_i^r} 
y(z) dz -\frac{1}{4}
\sum_{I=1}^{2N_f}
\int_{\widetilde{q}_I}^{\widetilde{\La}_0} 
y(z) dz + \frac{1}{2} \left(2N_c-4-2N_f \right) W(\La_0)
\nonu \\
&& + \frac{1}{2} \sum_{I=1}^{2N_f} W(z_I) - \pi i \left( 
2N_c-4-2N_f \right) S + 2\pi i \tau_0 S + 2\pi i 
\sum_{i=1}^{n} b_i S_i
\nonu
\eea
where
$S=S_0 + 2 \sum_{i=1}^{n} S_i $ and $z_I$ is the root of
\bea
B(z) = \mbox{det} \; m(z)=
\prod_{I=1}^{N_f}\left(z^2-z_I^2 \right)~.
\nonu
\eea

Since the curve (\ref{soy}) is same as the one 
(\ref{matrix-curve}) of $U(N_c)$ gauge
theory, we can use the integral results given there to
write down the effective superpotential as
\bea
W_{\rm eff} & = & 
 S  \left[  \frac{(N_c-2)}{2}  + \log 
\left( \frac{2^{\frac{N_c-2}{2}} m_A^{\frac{N_c-2}{2}} \La^{N_c-2-N_f}
    \mbox{det} z}{S^{\frac{N_c-2}{2}}} \right)
\right]
 \nonu \\
&& -S \sum_{I=1, r_I=0}^{N_f} \left[- \log 
\left( \frac{1}{2} + \frac{1}{2} \sqrt{1-
\frac{2S}{m_A z_I^2}} \right) + \frac{m_A z_I^2}{2S} \left( \sqrt{1-
\frac{2S}{m_A z_I^2}} -1 \right) + \frac{1}{2} \right] 
\nonu \\
&& -S \sum_{I=1, r_I=1}^{N_f} \left[- \log 
\left( \frac{1}{2} - \frac{1}{2} \sqrt{1-
\frac{2S}{m_A z_I^2}} \right) + \frac{m_A z_I^2}{2S} \left( -\sqrt{1-
\frac{2S}{m_A z_I^2}} -1 \right) + \frac{1}{2} \right] .
\nonu 
\eea

We can solve $M(z)$ and $T(z)$ as did for $U(N_c)$ gauge theory.
For simplicity we take all $r_I=0$, i.e., all poles at the 
second sheet. 
For the $I$-th block diagonal matrix element of $M(z)$
($I=1, \cdots, N_f$)  it is given by  
\bea
M_I(z) 
& =&  \left(\begin{array}{cc} 0 
& -\frac{R(z) -R(q_I=m_I)}{z-m_I} \\ \frac{R(z) - R(q_I=-m_I)}{z+m_I} & 0 
\end{array} \right)
\nonu 
\eea
where 
\bea
R(z) =m_A  \left(z -  \sqrt{z^2- 4\mu^2} \right). 
\nonu
\eea
Expanding $M_I(z)$ in the series of $z$ we can find 
$
\braket  { Q^f \Phi Q^{f'}
J_{ff'} + 
Q^f \widetilde{m}_{ff'} Q^{f'}} = 2 N_f S$.

The gauge invariant operator $T(z)$ can be constructed similarly
as follows:
\bea
T(z) = \frac{B^{\prime}(z)}{2B(z)}-\sum_{I=1}^{N_f} 
\frac{y(q_I) z_I }{ y(z) \left(z^2-z_I^2 \right)} +
\frac{c(z)}{y(z)} -\frac{2}{z} 
\frac{R(z)}{y(z)}
\label{sotz}
\eea
where
\bea
c(z) = \left< \Tr \frac{W^{\prime}(z) -W^{\prime}(\Phi)}{z-\Phi} 
\right> - \sum_{I=1}^{N_f} \frac{z W^{\prime}(z)-z_I W^{\prime}(z_I)}{
\left(z^2-z_I^2\right)}.
\nonu
\eea
For the theory without the quarks, 
the Konishi anomaly was derived in \cite{ac,krs,ao9}.
The last term in (\ref{sotz}) reflects the action of orientifold.
For our example we have 
\bea
c(z)= m_A \left( N_c-N_f \right). 
\nonu
\eea
and 
\bean
T(z) & = &  
\frac{1}{z} N_c +
\frac{1}{z^3}
\left[ \sum_{I=1}^{N_f}  z_I \left( z_I-  
\sqrt{z_I^2 - 4 \mu^2} \right)
+   2\mu^2 \left( N_c-2-N_f \right)   \right] \nonu \\
& & +
\frac{1}{z^5}
\left[ \sum_{I=1}^{N_f} z_I^4 -\sum_{I=1}^{N_f} \sqrt{z_I^2-4\mu^2}
  z_I 
\left( z_I^2 + 2\mu^2 \right) +
6\mu^4 \left( N_c-N_f \right) -12 \mu^4  \right] 
  + {\cal O} \left(\frac{1}{z^7} \right)
\nonu \\
\eean
where for equal mass of flavor, we get
$
\left< \Tr\, \Phi^2 \right> = N_f  q \left( q-  \sqrt{q^2 - 4 \mu^2} \right)
+   2\mu^2 \left( N_c-2-N_f \right)  
$
and
$
\left< \Tr\, \Phi^4 \right> = N_f q^4 -N_f \sqrt{q^2-4\mu^2} q 
\left( q^2 + 2\mu^2 \right) +
6\mu^4 \left( N_c-2-N_f \right)$.



Let us assume the mass of flavors are the same and $K$ of them
(in this case, $r_I=1$)
locate at the first sheet while the remainder $(N_f-K)$ where
$r_I=0$ are
at the second sheet. Then from the effective superpotential,  
it is ready to extremize this with respect to 
the glueball field $S$ 
\footnote{
One can easily check that this equation with parameters 
$(N_c, N_f, K)$ is equivalent to the one with 
parameters $(N_c-2r, N_f-2r, K-r)$. The equation of 
motion for glueball field is the same.
Since the equation of motion for both $r$-th Higgs branch and
$\left(N_f-r\right)$-th Higgs branch is the same, one expects that
both branches have some relation.
By redefinition of
$
S \rightarrow \frac{4m_A^2 \La^4}{S} \equiv \widetilde{S}, 
z_f \rightarrow \frac{2m_A
  \La^2}{S} z_f \equiv \widetilde{z_f}$
we get the final relation between $K$ Higgs branch and 
$\left(N_f-K \right)$ 
Higgs branch.}
\bea
0&=& \log \left( \frac{\La_1^{\frac{3}{2}(N_c-2)-N_f}}
{S^{\frac{N_c-2}{2}}} \right) \nonu\\
&&+ K \log\left(\frac{z_f-\sqrt{{z_f}^2-\frac{2S}{m_A}}}{2} \right) +
\left(N_f-K \right) \log \left( \frac{z_f+\sqrt{{z_f}^2-\frac{2S}{m_A}}}{2} \right).
\label{minimum}
\eea
Or by rescaling the fields 
$\Sh = \frac{S}{2m_A \La^2}$, $\zh_f = \frac{z_f}{\La} $
one gets the solution and consider for $K=0$ case  
\bea
1= \Sh^{-\frac{N_c-2}{2}} 
\left(  \frac{\zh_f+ \sqrt{\zh_f^2-4\Sh}}{2}  \right)^{N_f}.
\nonu
\eea
This equation is the same as the one 
in $U(N_c)$ case with $N_c\to \frac{N_c-2}{ 2}$, 
so the discussion of passing poles will go through without
modification and the result is  when $N_f\geq \frac{N_c-2}{ 2}$,
the on-shell 
poles at the second plane cannot pass the cut to
reach the first sheet far away from the cut.

By using the condition (\ref{minimum}), one gets the on-shell 
effective
superpotential
\bea
W_{\rm eff,on-shell}  =     \frac{1}{2} \left(N_c-2-N_f \right) S  
+ \frac{1}{2}  m_A {z_f}^2 \left( N_f 
 + \left( 2 K-N_f \right)   \sqrt{1-
\frac{2S}{m_A {z_f}^2}}  \right).  
\nonu 
\eea
It can be checked that this is the same as 
$ \frac{1}{2} m_A \left< \Tr\,  \Phi^2 \right>$.

\subsection{$USp(2N_c)$ theory with flavors}

\indent

For the $USp(2N_c)$ gauge theory with $N_f$ flavors we will sketch
the discussion because most of them 
is similar to $SO(N_c)$ gauge theory.
The tree level superpotential is given by (\ref{tree}) but with
 $J$ the symplectic metric, and $\widetilde{m}_{ff'}$
 the quark mass  given by 
\bea
J=\left(\begin{array}{cc} 0 & 1 \\ -1 & 0 \end{array} \right)
\otimes {\bf I}_{N_c \times N_c}, \qquad
\widetilde{m}=\left(\begin{array}{cc} 0 & -1 \\ 1 & 0 
\end{array} \right)
\otimes \mbox{diag} (m_1, \cdots, m_{N_f})~.
\nonu
\eea 
We parametrize   the matrix model curve and the resolvent $R(z)$ 
as
\bea
y_m^2 & = &  W^{\prime}(z)^2 + f(z) = m_A^2 \left( z^2 + 4 \mu^2 
\right), 
\nonu \\
R(z)& = & m_A  \left(z -  \sqrt{z^2+ 4\mu^2} \right). 
\nonu
\eea
The effective superpotential is given by
\bea
W_{\rm eff} & = & S   \left(N_c+1 \right) \left[ 1 + \log 
\left( \frac{\tilde{\La}^3}{S} \right)
\right] \nonu \\
&& -S \sum_{I=1, r_I=0}^{N_f} \left[- \log 
\left( \frac{1}{2} + \frac{1}{2} \sqrt{1+
\frac{2S}{m_A z_I^2}} \right) - \frac{m_A z_I^2}{2S} \left( \sqrt{1+
\frac{2S}{m_A z_I^2}} -1 \right) + \frac{1}{2} \right] 
\nonu \\
&& -S \sum_{I=1, r_I=1}^{N_f} \left[- \log 
\left( \frac{1}{2} - \frac{1}{2} \sqrt{1+
\frac{2S}{m_A z_I^2}} \right) - \frac{m_A z_I^2}{2S} \left( -\sqrt{1+
\frac{2S}{m_A z_I^2}} -1 \right) + \frac{1}{2} \right] 
\nonu 
\eea
and $T(z)$ is given by 
\bea
T(z) = \frac{B^{\prime}(z)}{2B(z)}-\sum_{I=1}^{N_f} 
\frac{y(q_I) z_I }{ y(z) \left(z^2-z_I^2 \right)} +
\frac{c(z)}{y(z)} +\frac{2}{z} 
\frac{R(z)}{y(z)}.
\nonu
\eea
Note the last term (different sign) compared with the $SO(N_c)$
gauge theory. From the solution of $M(z)$, we can show that
although
$
\left<  Q^f (\Phi J)_{ff'} Q^{f'}
+ 
Q^f \widetilde{m}_{ff'} Q^{f'} J 
 \right> = 2 N_f S\neq 0$, we still have
on-shell relation  
$ \frac{1}{2} m_A \left< \Tr\,  \Phi^2 \right> 
= W_{\rm eff}$.

The equation of motion is given by \footnote{
One can easily check that this equation with parameters 
$(2N_c, N_f, K)$ is equivalent to the one with 
parameters $(2N_c-2r, N_f-2r, K-r)$. In other words, the equation of 
motion for glueball field is the same.
Since the equation of motion for both $r$-th Higgs branch and
$\left(K-r\right)$-th Higgs branch is equivalent to each other, 
one expects that
both branches have some relation.} 
\bea
0= \log \Sh^{-N_c-1} + K \log 
\left(\frac{\zh_f-\sqrt{\zh_f^2+4\Sh}}{2} \right)
+\left( N_f-K \right) \log 
\left( \frac{\zh_f+\sqrt{\zh_f^2+4\Sh}}{2} \right)
\nonu
\eea
where $\Sh = \frac{S}{2m_A \La^2}$, $ \zh_f = \frac{z_f}{\La} 
$. From this we can read out the following result:
when $N_f\geq N_c+1$,
the on-shell poles at the second sheet cannot pass through 
the cut  to reach 
the first sheet far away from the cut.

\section*{Acknowledgments}

\indent

We would like to thank Freddy Cachazo, Eric D'Hoker, Ken
Intriligator, Romuald Janik, Per Kraus, and Hirosi Ooguri for
enlightening discussions.  This research of CA was supported by a grant
in aid from the Monell Foundation through Institute for Advanced Study,
by SBS Foundation, and by Korea Research Foundation Grant
(KRF-2002-015-CS0006).  The work of BF is supported by the Institute for
Advanced Study under NSF grant PHY-0070928.  The work of YO was
supported by JSPS Research Fellowships for Young Scientists.  The work
of MS was supported by NSF grant 0099590.



\appendix


\renewcommand{\theequation}{\Alph{section}\mbox{.}\arabic{equation}}

\bigskip\bigskip
\noindent
{\LARGE \bf Appendix}

\section{\large \bf On matrix model curve with $N_f(>N_c)$  
flavors  
}
\setcounter{equation}{0}

\indent

In this Appendix we prove by strong coupling analysis that matrix model
curve corresponding to $U(N_c)$ supersymmetric gauge theory with $N_f$
flavors is exactly the same as the one without flavors when the degree
$(k+1)$ of tree level superpotential $W_{\rm tree}$ is less than 
$N_c$
\footnote{The generalized Konishi anomaly equation of $R(z)$ given in
(\ref{csw2-2.14}) is same with or without flavors, so the form of the
solution is the same for gauge theory with or without flavor.  In this
Appendix we use another method to prove this result.}.  This was first
proved in \cite{Ookouchi} but the derivation was valid only for the
range $N_f<N_c$. Then in \cite{bfhn}, the proof was extended to the
cases with the range $2N_c > N_f\ge N_c$.  However, in \cite{bfhn}, the
characteristic function $P_{N_c}(x)$ was defined by $P_{N_c}(x)=\det
(x-\Phi)$, without taking into account the possible quantum corrections
due to flavors.  In consequence, it appeared that the matrix model
curve is changed by addition of flavors. In this Appendix, we use the
definition of $P_{N_c}(x)$ proposed in Eq.\ (C.2) of \cite{CSW-II}:
\begin{eqnarray}
P_{N_c}(x)=x^{N_c} \mbox{exp}\left(-\sum_{i=1}^{\infty}\frac{U_i}{x^i} 
\right)+\La^{2N_c-N_f}\frac{\widehat{B}(x)}{x^{N_c}}\mbox{exp}\left(
\sum_{i=1}^{\infty}\frac{U_i}{x^i} \right), \label{PN2}
\end{eqnarray}
which incorporates quantum corrections and reduces to $P_{N_c}(x)=\det
(x-\Phi)$ for $N_f=0$, and see clearly that the matrix model curve is
not changed, even when the number of flavors is more than $N_c$.  
Since
$P_{N_c}(x)$ is a polynomial in $x$, (\ref{PN2}) can be used to express
$U_r$ with $r>N_c$ in terms of $U_r$ with $r\le N_c$ by imposing the
vanishing of the negative power terms in $x$.

Assuming that the unbroken gauge group at low energy is $U(1)^n$ with
$n\le k$, the factorization form of Seiberg--Witten curve can be written
as,
\begin{eqnarray}
P_{N_c}^2(x)-4\La^{2N_c-N_f}\widehat{B}(x)=H_{N_c-n}^2(x)F_{2n}(x).
\nonu
\end{eqnarray}
The effective superpotential with this double root constraint can 
be written as follows
\footnote{If we want to generalize this proof 
to more general cases in which $k+1$ is greater than and equals 
to $N_c$, we have to take care more constraints like Appendix A in
\cite{csw1}, which should be straightforward.}. 
\begin{eqnarray}
W_{\rm eff} & = & \sum_{r=0}^{k}g_rU_{r+1}
\nonu \\
& + & 
\sum_{i=1}^{N_c-n}\left(L_i \oint \frac{P_{N_c}(x)-2\epsilon_i 
\La^{N_c-\frac{N_f}{2}}\sqrt{\widehat{B}(x)}}{x-p_i}dx+B_i \oint 
\frac{P_{N_c}(x)-2\epsilon_i \La^{N_c-\frac{N_f}{2}}\sqrt{\widehat{B}
(x)}}{(x-p_i)^2}dx \right).
 \nonu
\end{eqnarray}
The equations of motion for $B_i$ and $p_i$ are given as follows
respectively:
\begin{eqnarray}
0=\oint \frac{P_{N_c}(x)-2\epsilon_i \La^{N_c-\frac{N_f}{2}}\sqrt
{\widehat{B}(x)}}{(x-p_i)^2}dx,\quad 0=2B_i \oint \frac
{P_{N_c}(x)-2\epsilon_i \La^{N_c-\frac{N_f}{2}}\sqrt{\widehat{B}(x)}}
{(x-p_i)^3}dx. 
\nonu
\end{eqnarray}
Assuming that the factorization form does not have any triple 
or higher
roots, we obtain $B_i=0$ at the level of equation of motion. 
Next we
consider the equation of motion for $U_r$:
\begin{eqnarray}
0=g_{r-1}+\sum_{i=1}^{N_c-n}\oint \left[\frac{P_{N_c}}{x^r}-2
\frac{x^{N_c}}{x^r}\mbox{exp}\left( -\sum_{i=1}^{\infty} 
\frac{U_i}{x^i} \right) \right]\frac{L_i}{x-p_i}dx
\nonu
\end{eqnarray}
where we used $B_i=0$ and (\ref{PN2}) to evaluate $\frac{\partial
P_{N_c}}{\partial U_r}$. Now, as in \cite{csw1}, we multiply this by
$z^{r-1}$ and sum over $r$.
\begin{eqnarray}
W^{\prime}(z)=-\oint \frac{P_{N_c}}{x-z}\sum_{i=1}^{N_c-n} \frac{L_i}
{x-p_i}dx+\oint \frac{2x^{N_c}}{x-z}\mbox{exp} \left(-\sum_{k=1}^{
\infty}\frac{U_k}{x^k} \right)\sum_{i=1}^{N_c-n}\frac{L_i}{x-p_i}dx. 
\nonu
\end{eqnarray}
Defining the polynomial $Q(x)$ in terms of 
\begin{eqnarray}
\sum_{i=1}^{N_c-n}\frac{L_i}{x-p_i}=\frac{Q(x)}{H_{N_c-n}(x)},
\label{Qpolynomial}
\end{eqnarray}
and also using (\ref{PN2}) and factorization form, we obtain
\begin{eqnarray}
W^{\prime}(z)&=&-\oint \frac{P_{N_c}}{x-z}\frac{Q(x)}{H_{N_c-n}(x)} 
dx+\oint \frac{P_{N_c}}{x-z}\frac{Q(x)}{H_{N_c-n}(x)} dx +\oint 
\frac{Q(x)\sqrt{F_{2n}(x)}}{x-z} dx \nonu \\
&=& \oint \frac{Q(x)\sqrt{F_{2n}(x)}}{x-z} dx.
\nonu
\end{eqnarray}
This is nothing but $(2.37)$ in \cite{csw1}. Since 
$W^{\prime}(z)$ is a polynomial of degree $k$, the $Q$ 
should be a polynomial of degree $(k-n)$. Therefore, we 
conclude that the matrix model curve is not changed by 
addition of flavors:
\begin{eqnarray}
y_m^2=F_{2n}(x)Q^2_{k-n}(x)=W_k^{\prime}(x)^2+{\cal O}(x^{k-1}). 
\label{Nocorrection}
\end{eqnarray}

\section{\large \bf Equivalence between $W_{\rm low}$ and 
$W_{\rm eff}(\ev{S_i})$ with flavors \label{CVmethod}} 
\setcounter{equation}{0}

\indent

In this Appendix we prove the equivalence $W_{\rm low}$ in $U(N_c)$
gauge theory with $W_{\rm eff}(\langle S_i \rangle)$ in corresponding
dual geometry when some of the branch cuts on the Riemann surface are
closed and the degree $(k+1)$ of the tree level superpotential $W_{\rm
tree}$ is less than $N_c$. This was first proved in \cite{Ookouchi},
however, the proof was only applicable in the $N_f<N_c$ cases.
Especially, the field theory analysis in \cite{Ookouchi} did not work
for $N_c \le N_f< 2N_c$ cases. 
Furthermore, as we saw in the main text, for some particular choices of
$z_I$ (position of the flavor poles), extra double roots appear in the
factorization problem.  In section \ref{sec:2cut}, we dealt with $U(3)$
with cubic tree level superpotential and saw the equivalence of two
effective superpotentials for such special situations. To include these
cases we are interested in the Riemann surface that has some closed
branch cuts. Therefore our proof is applicable for $U(N_c)$ gauge
theories with $W_{\rm tree}$ of degree $k+1$ $(<N_c)$ in which some of
branch cuts are closed and number of flavors is in the range $N_c\le N_f
< 2N_c$. In addition, we restrict our discussion to the Coulomb phase.

In the discussion below, we follow the strategy developed by 
Cachazo and Vafa in \cite{CV} and use (\ref{PN2}) as the 
definition of $P_{N_c}(x)$. We have only to show the two
relations:
\begin{eqnarray}
W_{\rm low}(g_r,z_I,\La)\big|_{\La \to 0} &=&W_{\rm eff}
(\langle S_i \rangle)\big|_{\La \to 0}, \label{proof1} \\
\frac{\partial W_{\rm low}(g_r,z_I,\La )}{\partial \La}&=&
\frac{\partial W_{\rm eff}(\langle S_i \rangle )}{\partial 
\La}, \label{proof2}
\end{eqnarray}
the equivalence of two effective superpotentials in the classical limit
and that of the derivatives of the superpotentials with respect to
$\La$.

\subsection{Field theory analysis }

\indent

Let $k$ be the order of $W_{\rm tree}^{\prime}$ and 
$n$$(\le k)$ be the number of
$U(1)$ at low energy. Since we are interested in cases with 
degenerate branch cuts, let us consider the following factorization 
form
\footnote{In the computation below, we will use relation 
(\ref{Qpolynomial}) and put $g_{k+1}=1$.}:
\begin{eqnarray}
P_{N_c}(x)^2-4\La^{2N_c-N_f}\widehat{B}(x)&=
&F_{2n}(x)\left[Q_{k-n}(x)
\widetilde{H}_{N_c-k}(x)\right]^2 \equiv F_{2n}(x)\left[H_{N_c-n}(x)
\right]^2, \label{FactoForm} \\
W^{\prime}(x)^2+f_{k-1}(x)&=&F_{2n}(x)Q_{k-n}(x)^2 .
\nonu
\end{eqnarray}
If $k$ equals to $n$, all the branch cuts in $F_{2k}(x)$ are open. 
The low energy effective superpotential is given by
\begin{eqnarray}
&& W_{\rm low}  = 
\sum_{r=1}^{n+1}g_r U_r \nonu \\
&&+\sum_{i=1}^l 
\left[L_i\left(P_{N_c}(p_i)-2\epsilon_i 
\La^{N_c-\frac{N_f}{2}}\sqrt{\widehat{B}(p_i)} \right)  +
Q_i \frac{\partial}{\partial p_i}
\left(P_{N_c}(p_i)-2\epsilon_i 
\La^{N_c-\frac{N_f}{2}}\sqrt{\widehat{B}(p_i)} \right) \right], 
\nonu 
\end{eqnarray}
where $l\equiv N-n$ and $P_{N_c}(x)$ is defined by (C.3) or (C.4)
in \cite{CSW-II},
\begin{eqnarray}
P_{N_c}(x)=\ev{ \det (x-\Phi) } +\left[ \La^{2N_c-N_f} 
\frac{\widehat{B}(x)}{x^{N_c}} \mbox{exp} \left( \sum_{i=1}^{\infty} 
\frac{U_i}{x^i} \right)  \right]_+. \label{rr1}
\end{eqnarray}
The second term is specific to the $N_f \ge N_c$ case, 
representing quantum correction.  Define
\begin{eqnarray}
K(x)\equiv  \left[ \La^{2N_c-N_f} \frac{\widehat{B}(x)}{x^{N_c}} 
\mbox{exp} \left( \sum_{i=1}^{\infty} \frac{U_i}{x^i} \right)  
\right]_+. 
\nonu
\end{eqnarray}
The first term in (\ref{rr1}) can be represented as  $\langle 
\det (x-\Phi)   \rangle \equiv \sum_{k=0}^{N_c}x^{N_c-k}s_k$. The 
relation between $U_i$'s and $s_k$'s are given by the ordinary 
Newton
relation, $ks_k+\sum_{r=1}^krU_rs_{k-s}=0$. 
From the variations of $W_{\rm low}$ with respect to $p_i$ and 
$Q_i$, we
conclude that $Q_i=0$ at the level of the equation of motion. 
In
addition, the variation of $W_{\rm low}$ with respect to $U_r$ 
leads to
\begin{eqnarray}
g_r&=&-\sum_{i=1}^lL_i \frac{\partial P_{N_c}(p_i)}{\partial U_r} 
=\sum_{i=1}^l\sum_{j=0}^{N_c} L_i p_i^{N_c-j}s_{j-r}-\sum_{i=1}^l 
L_i \frac{\partial K(p_i)}{\partial U_r} \nonu \\
&=&\sum_{i=1}^l\sum_{j=0}^{N_c} L_i p_i^{{N_c}-j}s_{j-r}-\sum_{i=1}^l 
L_i \left[ \La^{2{N_c}-N_f}\frac{\widehat{B}(p_i)}{p_i^{{N_c}+r}} 
\mbox{exp} \left(\sum_{k=1}^{\infty} \frac{U_k}{p_i^k} \right) 
\right]_+ .
\nonu
\end{eqnarray}
Let us define
\begin{eqnarray}
{\cal G}_r(p_i) \equiv \left[ \La^{2{N_c}-N_f}\frac{
\widehat{B}(p_i)}{p_i^{{N_c}+r}} \mbox{exp} \left(\sum_{k=1}^
{\infty} \frac{U_k}{p_i^k} \right) \right]_+.
\nonu
\end{eqnarray}
By using these relations, let us compute $W_{\rm cl}^{\prime}$
\begin{eqnarray}
W_{\rm cl}^{\prime}&=&\sum_{r=1}^{N_c} g_r x^{r-1}\nonu \\
&=& \sum_{r=-\infty}^{N_c} \sum_{i=1}^l \sum_{j=0}^{N_c} x^{r-1}p_i^
{{N_c}-j}s_{j-r}L_i -\frac{1}{x} \sum_{i=1}^l L_i \det (p_i-\Phi)-
\sum_{r=1}^{{N_c}}\sum_{i=1}^{l}L_i {\cal G}_{r}(p_i)x^{r-1}
\nonu \\
&=& \sum_{i=1}^l \frac{\det (x-\Phi)}{x-p_i}L_i -\frac{1}{x} 
\sum_{i=1}^l L_i \det (p_i-\Phi)-\sum_{r=1}^{{N_c}}\sum_{i=1}^{l}L_i 
{\cal G}_{r}(p_i)x^{r-1}
\nonu \\
&=& \sum_{i=1}^l \frac{P_{N_c}(x)}{x-p_i}L_i-\sum_{i=1}^l \frac{K(x)}
{x-p_i}L_i-\frac{1}{x}\sum_{i=1}^lL_i P_{N_c}(p_i)+\frac{1}{x}
\sum_{i=1}^l L_i K(p_i)
\nonu \\
&&  -\sum_{r=1}^{N_c}\sum_{i=1}^l L_i 
{\cal G}_{r}(p_i)x^{r-1} 
\label{rr2} 
\end{eqnarray}
where we dropped $ {\cal O}(x^{-2})$.
The fifth term above can be written as
\begin{eqnarray}
-\sum_{r=1}^{N_c}\sum_{i=1}^l L_i {\cal G}_{r}(p_i)x^{r-1}= -
\sum_{r=-\infty}^{N_c}\sum_{i=1}^l L_i {\cal G}_{r}(p_i)x^{r-1}+
\sum_{i=1}^l \frac{1}{x} L_i K(p_i)+{\cal O}(x^{-2}).
\nonu
\end{eqnarray}
After some manipulation with the factorization form 
(\ref{FactoForm}), we obtain a relation
\begin{eqnarray}
Q_{k-n}(x)\sqrt{F_{2n}(x)} &=& \frac{P_{N_c}(x) }{\widetilde{H}_{{N_c}-k}}- 
\frac{2\La^{2{N_c}-N_f}\widehat{B}(x)}{\widetilde{H}_{{N_c}-k}x^N}\mbox{exp} 
\left(\sum_{i=1}^{\infty} \frac{U_i}{x^i} \right)\nonu \\
&=& \frac{P_{N_c}(x) }{\widetilde{H}_{{N_c}-k}}-\frac{2K(x)}{
\widetilde{H}_{{N_c}-k}}+{\cal O}(x^{-2}). 
\nonu
\end{eqnarray}
Substituting this relation into (\ref{rr2}) we obtain 
\begin{eqnarray}
W_{\rm cl}^{\prime} & = & 
Q_{k-n}(x)\sqrt{F_{2n}(x)}-\sum_{i=1}^l \frac{1}{x} 
\left[L_iP_{N_c}(p_i)-2L_i K(p_i) \right]+\sum_{i=1}^l \frac{K(x)}
{x-p_i}L_i\nonu \\
&&-\sum_{r=-\infty}^{N_c} \sum_{i=1}^l L_i {\cal G}_r(p_i)x^{r-1}+
{\cal O}(x^{-2}).  
\nonu
\end{eqnarray}
Finally let us use the following relation
\footnote{
For simplicity, we ignore $\sum_{i=1}^{l}L_i$. To prove the 
relation, let us consider the circle integral over $C$. Until 
now, since we assumed $p_i<x$, the point $p_i$ should be 
included in the contour $C$. Multiplying $\frac{1}{x^k}$ $k\ge 0$ 
and taking the circle integral, we can pick up the coefficient 
$c_{k-1}$ of $x^{k-1}$. In addition, if we denote a polynomial 
$M\equiv \La^{2{N_c}-N_f}\frac{\widehat{B}(x)}{x^{N_c}}\mbox{exp}
\left(\sum_{i=1}^{\infty}\frac{U_i}{x^i} \right)\equiv \sum_
{j=-\infty}^{\infty}a_jx^j$
we can obtain following relation,
\begin{eqnarray}
\frac{\left[M \right]_+}{x^k}=\left[ \frac{M}{x^k} \right]_+ 
+\sum_{j=0}^{k-1}a_j x^{-(k-j)} \iff \frac{K(x)}{x^k}={\cal 
G}_k(x) +\sum_{j=0}^{k-1}a_j x^{-(k-j)}\nonu
\end{eqnarray}
where right hand side means circle integral of left hand side. 
Thus, we obtain the $c_{k-1}$ as
\begin{eqnarray}
c_{k-1}= \sum_{j=0}^{k-1} \oint_{x=0}\frac{a_j x^{j-k}}{x-p_i}dx 
+\sum_{j=0}^{k-1} \oint_{x=p_i}\frac{a_j x^{j-k}}{x-p_i}dx 
= -\sum_{j=0}^{k-1}\oint_{x=0} a_j x^{j-k}\sum_{n=1}^{\infty}
\frac{x^n}{p_i^{n+1}} dx +\sum_{j=0}^{k-1} a_j p_i^{j-k}=0 \nonu
\end{eqnarray}
where we used that around $x=0$,
$\frac{1}{x-p_i}=-\sum_{n=1}^{\infty}\frac{x^n}{p_i^{n+1}}$. 
Therefore
the left hand side of (\ref{mar67}) can be written as 
$\sum_{j=-\infty}^{\infty}c_{j}x^j=\sum_{j=-\infty}^{-2}c_jx^j=
{\cal O}(x^{-2})$.
}
;
\begin{eqnarray}
\sum_{i=1}^l \frac{K(x)}{x-p_i}L_i-\sum_{r=-\infty}^{N_c} 
\sum_{i=1}^l L_i {\cal G}_r(p_i)x^{r-1}={\cal O}(x^{-2}). 
\label{mar67}
\end{eqnarray}

After all, by squaring $W_{\rm cl}^{\prime}$, we have 
\begin{eqnarray}
Q_{k-n}(x)^2F_{2n}(x)&=&W_{\rm cl}^{\prime 2}(x)+ 2
\sum_{i=1}^l \frac{1}{x} \left[L_iP_{N_c}(p_i)-2L_i K(p_i) 
\right]x^{k-1} +{\cal O}(x^{k-2}), \nonu \\
b_{k-1}&=& 2\sum_{i=1}^l \frac{1}{x} \left[L_iP_{N_c}(p_i)-
2L_i K(p_i) \right].
\nonu
\end{eqnarray}

On the other hand the variation of $W_{\rm low }$ with respect to $\La$
is given by
\begin{eqnarray}
\frac{\partial W_{\rm low}}{\partial \log \La^{2{N_c}-N_f} }=
\sum_{i=1}^lL_iK(p_i)-\frac{1}{2}\sum_{i=1}^lL_i P_{N_c}(p_i)=-
\frac{b_{k-1}}{4} \label{db_{n-1}}.
\end{eqnarray}
This is one of the main results for our proof.  In the dual geometry
analysis below, we will see the similar relation.

In the classical limit, we have only to consider the expectation value
of $\Phi$. In our assumption, gauge symmetry breaks as $U({N_c})\to
\prod_i^{n} U(N_{c,i})$ we have $\Tr\,\Phi=\sum_{i=1}^n
N_{c,i}a_i$. Therefore in the classical limit $W_{\rm low}$ behaves as
\begin{eqnarray}
W_{\rm cl}=\sum_{i}^n N_{c,i} W(a_i). \label{Apr1}
\end{eqnarray}
By comparing (\ref{db_{n-1}}) and (\ref{Apr1}) and the 
similar result which we will see in the dual geometry 
analysis below, we will show (\ref{proof1}) and (\ref{proof2}). 
Let us move to the dual geometry analysis.

\subsection{Dual geometry analysis with some closed branch cuts}

\indent

As we have already seen in the main text, a solution with $\langle S_i
\rangle=0$ appears for some special choice of $z_I$, the position of
flavor poles.
In our present proof, however, we put some of $S_i$ to be zero from the
beginning, {\em without specifying $z_I$\/}. More precisely, what we
prove in this Appendix is as follows: for a given choice of $z_I$, if
there exists a solution to the factorization problem with some of
$\langle S_i \rangle$ vanishing, then we can construct a dual geometry
which gives the same low energy effective superpotential as the one
given by the solution to the factorization problem, by setting some of
$S_i$ to zero from the beginning. Therefore, this analysis {\em does
not\/} tell us when a solution with $\langle S_i \rangle=0$ appears. To
know that within the matrix model formalism, we have to go back to
string theory and consider an explanation such as the one given in
\cite{IKRSV}.

Now let us start our proof. Again, let $k$ be the degree of tree level
superpotential $W_{\rm tree}^{\prime}(x)$ and $n$ be the number of
$U(1)$ at low energy.  To realize this situation, we need to consider
that $(k-n)$ branch cuts on the Riemann surface should be closed, which
corresponds to $\langle S_i \rangle=0$. Here, there is one important
thing: As we know from the expansion of $W_{\rm eff}$ in terms of $\La$
(e.g. see (\ref{juho8Apr04})), we cannot obtain any solutions with
$\langle S_i \rangle=0$ if we assume that $S_i$ is dynamical and solve
its equation of motion.  Therefore to realize the situation with
vanishing $\langle S_i \rangle$, we must put $S_i=0$ at the off-shell
level by hand. With this in mind, let us study dual geometry which
corresponds to the gauge theories above. In the field theory, we assumed
that the Riemann surface had $(k-n)$ closed branch cuts. Thus, in this
dual geometry analysis we must assume that at off-shell level, $(k-n)$
$S_i$'s must be zero. For convenience, we assume that first $n$ $S_i$'s
are non-zero and the remaining $(k-n)$ vanish,
\begin{eqnarray}
S_i\neq 0 ,\ \ i=1,\cdots n,\ \ \ S_i=0,\ \  i=n+1,\cdots k.
\nonu
\end{eqnarray}
Therefore the Riemann surface can be written as
\begin{eqnarray}
y^2=F_{2n}(x)Q_{k-n}^2 =W^{\prime}(x)^2+b_{k-1}x^{k-1}+\cdots
\nonu
\end{eqnarray}
The effective superpotential in dual geometry corresponding to
$U({N_c})$ gauge theory with $N_f$ flavors was given in \cite{CSW-II}
(See also (\ref{Wexact-y})) and in the classical limit it behaves as
\footnote{Remember that in this Appendix we are assuming only Coulomb
branch. For the Higgs branch, see (7.11) and (7.12) in \cite{CSW-II}.}
\begin{eqnarray}
W_{\rm eff} \big|_{\rm cl}= \sum_{i=1}^n N_{c,i}W(a_i). \label{sqb}
\end{eqnarray}

As discussed in the previous section, existence of flavors does not
change the Riemann surface $y(x)$. In other words, Riemann surface is
not singular at $x_I$ (roots of $B(x)$),
\begin{eqnarray}
\oint_{x_I} y(x) dx =0,\qquad y(x)=\sqrt{W^{\prime}(x)^2+
b_{k-1}x^{k-1}+\cdots}.
\nonu
\end{eqnarray}
Therefore as in \cite{CV}, by deforming contours of all $S_i$'s 
and evaluating the residue at infinity on the first sheet, we 
obtain the following relation, 
\begin{eqnarray}
\sum_{i=1}^{n}S_i=\sum_{i=1}^{k}S_i=-\frac{1}{4}b_{k-1},
\nonu
\end{eqnarray}
where we used $\sum_{i=n+1}^kS_i=0$. 

With this relation in mind, next we consider the variation 
of $W_{\rm eff}$ with respect to $S_i$: 
\begin{eqnarray}
\frac{\partial W_{\rm eff}(S_i,\La)}{\partial S_i}=0, \quad 
i=1\cdots n. \label{beom}
\end{eqnarray}
Solving these equations, we obtain the expectation values, $\langle S_i
\rangle$. Of course, these vacuum expectation values depend on $\La$, $g_r$ and $z_I$. Thus
when we evaluate the variation of $W_{\rm eff}(\langle S_i \rangle ,\La
)$ with respect to $\La$, we have to pay attention to implicit
dependence on $\La$. However the implicit dependence does not contribute
because of the equation of motion (\ref{beom}):
\begin{eqnarray}
\frac{d W_{\rm eff}(\langle S_i \rangle, \La )}{d \La }&=&
\sum_{i=1}^n \frac{ \partial \langle S_i \rangle }{\partial 
\La}\cdot \frac{\partial W_{\rm eff}(\langle S_i \rangle , 
\La ) }{\partial \langle S_i \rangle }+\frac{\partial W_{\rm 
eff}(\langle S_i \rangle , \La ) }{\partial  \La  } \nonu \\
&=&\frac{\partial W_{\rm eff}(\langle S_i \rangle , \La ) }
{\partial  \La  }.
\nonu
\end{eqnarray}
On the other hand, explicit dependence on $\La$ can be easily obtained
by monodromy analysis. Here let us recall the fact that the presence of
fundamentals does not change the Riemann surface.  In fact, looking at
(\ref{Wexact-y}) we can read off the dependence from the term $2\pi i
\tau_0= \log \left( \frac{B_L \La^{2{N_c}-N_f}}{\La_0^{2{N_c}-L}}
\right)$,
\begin{eqnarray}
\frac{d W_{\rm eff}(\langle S_i \rangle, \La )}{d \log 
\La^{2{N_c}-N_f} }=S=-\frac{b_{k-1}}{4}. \label{mar63}
\end{eqnarray}

To finish our proof, we have to pay attention to $f_{k-1}(x)$,
on-shell. Namely putting $\langle S_i \rangle$ into $f_{k-1}(x)$ what
kind of property does it have? To see it, let us consider change of
variables from $S_i$'s to $b_i$'s.  As discussed in \cite{CDSW} the
Jacobian of the change is non-singular if $0\le j \le k-2$,
\begin{eqnarray}
\frac{\partial S_i}{\partial b_j}=-\frac{1}{8\pi i}
\oint_{A_i}dx \frac{x^j}{\sqrt{W^{\prime}(x)^2+f(x)}}.
\nonu
\end{eqnarray}
In our present case, since only $n$ of $k$ $S_i$'s are 
dynamical variable, we use $b_i$, $i=0,\cdots n-1$ in a function 
$f_{k-1}(x)=\sum b_ix^i$ as new variables, instead of $S_i$'s. 
As discussed in \cite{CV,Ookouchi,CSW-II}, by using Abel's 
theorem, the equation of motions for $b_i$'s is interpreted 
as an existence condition of a meromorphic function that has an 
${N_c}$-th order pole at infinity on the first sheet and an $({N_c}-N_f)$-th 
order zero at infinity on the second sheet of $\Sigma$ and a 
first order zero at $\widetilde{q}_I$. For a theory with 
$N_f\le 2{N_c}$, such a function can be constructed as follows 
\cite{NSW02,CSW-II}:
\begin{eqnarray}
\psi (x)=P_{N_c}(x)+\sqrt{P^2_{N_c}(x)-4\La^{2{N_c}-N_f}\widehat{B}(x)}.
\nonu
\end{eqnarray}
For this function to be single valued on the matrix model curve 
$y(x)$, the following condition must be satisfied,
\begin{eqnarray}
P_{N_c}^2(x)-
4\La^{2{N_c}-N_f}\widehat{B}(x)&=&F_{2n}(x){H^2_{{N_c}-n}}(x) \nonu
\\
W^{\prime}(x)^2+f(x)&=&F_{2n}(x)Q^2_{k-n}(x)
\nonu
\end{eqnarray}
This is exactly the same as the factorization form we already 
see in the field theory analysis. Therefore the value 
$b_{k-1}$ of on-shell matrix model curve in dual theory is 
the same one for field theory analysis. Comparing two results, 
(\ref{db_{n-1}}) and (\ref{Apr1}) with
corresponding results for the dual geometry analysis, 
(\ref{sqb}) and
(\ref{mar63}) we have shown the equivalence between these
two descriptions of
effective superpotentials.

\section{Computation of superpotential --- gauge theory side}



\label{comp_spot-gt}

\indent

In this Appendix, we demonstrate the factorization method used in
subsection \ref{QuantumMassless} to compute the low energy
superpotential, taking the $N_f=4$ case as an example.  Therefore there
are two kinds of solutions for the factorization problem
\eqref{factorization} and \eqref{p3h1}.

\bigskip\noindent
$\bullet$ {\bf The breaking pattern $\widehat{U(2)}\times U(1)$}

\indent

The first kind of solution for the factorization problem is given by
 \begin{eqnarray}
A=0 ,\qquad
B =  2a,
 \qquad 
 C= a^2-4\La^2,
\qquad 
  D= 0,\qquad 
   c = 0,
\qquad
  b= 0.
\nonu
\end{eqnarray} 
In the classical limit $\Lambda\to 0$, we can see the 
characteristic function goes as $P_3(x)\to x^2 \left(x+a\right)$, 
which means that the breaking pattern is $\widehat{U(2)}
\times U(1)$.  Note that since we are assuming $m_f=0$,
the notation ``$~~\widehat{}~~$'' should be used for the gauge group
that corresponds to the cut near the critical point at $x=0$.
Inserting these solutions into (\ref{constraint}) we obtain 
one constraint,
\begin{eqnarray}
\Delta^2=a^2+8\La^2. 
\nonu
\end{eqnarray}
We can easily represent $a$ as a Taylor expansion of $\La$: 
\begin{eqnarray}
a=-1 + 4\,T + 8\,T^2 + 32\,T^3 + 160\,T^4 + 896\,T^5 + 5376\,
T^6+\cdots,
\nonu
\end{eqnarray}
where we put $\Delta=1$ and defined $T\equiv \La^2$.
\footnote{If we take care  of a phase factor of $\La$, 
we will obtain the effective superpotentials corresponding 
to each vacuum. However in our present calculation, we want 
to check whether the effective superpotentials of two method, 
field theory and dual geometry, agree with each other. 
Therefore, we have only to pay attention to the coefficients in 
$W_{\rm low}$, neglecting the phase factor.}
 The coefficients of $P_3(x)$ are related to the
Casimirs $U_j=\frac{1}{ j}\ev{\Tr[\Phi^j]}$ as follows.  
For $N_c=3$,
$N_f=4$, \eqref{PN2} reads
\begin{align}
 P_3(x)
 &=
 x^3 \exp\biggl(-\sum_{j=1}^{\infty}\frac{U_j}{x^j} \biggr)
 +\La^{5}\frac{x^{4}}{x^3}\exp\biggl(\sum_{j=1}^{\infty}
\frac{U_j}{x^j} \biggr)\cr
 &=
 x^3-U_1 x^2+\left(-U_2+\frac{U_1^2}{ 2}+\La^2 \right)x
 +\left(-U_3+U_1U_2-\frac{U_1^3}{6}-\La^2 U_1 \right)+\cdots.\ 
\nonu
\end{align}
Comparing the coefficients, we obtain
\begin{align}
 U_1=-a,\qquad
 U_2=-b+\frac{a^2}{ 2}+\La^2 ,\qquad
 U_3=-c+ab-\frac{a^3}{ 3}-a \La^2 .
\nonu
\end{align}
Furthermore, one can compute $a_{1,2}$ from \eqref{constraint}.
Plugging all these into \eqref{W_low_U(3)}, we finally obtain
\begin{eqnarray}
W_{\rm low}=W_{\rm cl} + T - 6T^2 - \frac{40T^3}{3} - 
         56T^4 - 288T^5 - \frac{4928T^6}{3}+
        \cdots,\quad
         T\equiv \Lambda^2,
 \quad W_{\rm cl} =-\frac{1}{6}.\nonu
\end{eqnarray}

\noindent
$\bullet$ {\bf The breaking pattern $\widehat{U(1)}\times U(2)$}

\indent

The other kind of solution for the factorization problem is given by
 \begin{eqnarray}
 A&=& \frac{1}{2}(-a-2\eta \La), \qquad 
B = a-2\eta \La ,
 \qquad 
 C= \frac{1}{4}(a+2\eta \La)^2, \nonu \\
&&
  D= 0,\qquad 
   c = 0,
\qquad
  b= \frac{1}{4}(a+2\eta \La)^2
\nonu
\end{eqnarray}
where $\eta \equiv \pm1$. These solutions correspond to the 
breaking pattern $\widehat{U(1)}\times U(2)$ in the classical 
limit. Inserting these solutions into (\ref{constraint}) we 
obtain one constraint,
\begin{eqnarray}
\Delta^2=\frac{1}{4}(a^2-20a\, \eta\, \La +4\La^2). 
\nonu
\end{eqnarray}
Again, let us represent $a$ as a Taylor series of $\La$: 
\begin{eqnarray}
a=-2 + 10\,T - 24\,T^2 + 144\,T^4 - 1728\,T^6+\cdots,
\nonu
\end{eqnarray}
where we put $\Delta=1$, $\eta=1$ and defined $T\equiv \La$.
Doing the same way as previous breaking pattern, 
we can compute the effective superpotential as 
\begin{eqnarray}
        W_{\rm low} =W_{\rm cl} + 2T - 13T^2 + \frac{176T^3}{3} - 
         138T^4 + 792T^6 +\cdots,\quad
         T\equiv \Lambda, \qquad W_{\rm
         cl}=-\frac{1}{3}.  \nonu
\end{eqnarray}

The other cases with $N_f=1,2,3$ and $5$ can be done analogously.



\begin{thebibliography}{99}

\bibitem{CSW-II}
F.~Cachazo, N.~Seiberg and E.~Witten,
``Chiral Rings and Phases of Supersymmetric Gauge Theories,''
JHEP {\bf 0304}, 018 (2003)
[arXiv:hep-th/0303207].

\bibitem{DV}
R.~Dijkgraaf and C.~Vafa,
``Matrix models, topological strings, and supersymmetric gauge theories,''
Nucl.\ Phys.\ B {\bf 644}, 3 (2002)
[arXiv:hep-th/0206255];
R.~Dijkgraaf and C.~Vafa,
``On geometry and matrix models,''
Nucl.\ Phys.\ B {\bf 644}, 21 (2002)
[arXiv:hep-th/0207106];
%
R.~Dijkgraaf and C.~Vafa,
``A perturbative window into non-perturbative physics,''
[arXiv:hep-th/0208048].

\bibitem{IKRSV}
K.~Intriligator, P.~Kraus, A.~V.~Ryzhov, M.~Shigemori and C.~Vafa,
``On low rank classical groups in string theory, gauge theory and matrix
models,''
Nucl.\ Phys.\ B {\bf 682}, 45 (2004)
[arXiv:hep-th/0311181].

\bibitem{Kraus:2003jf}
P.~Kraus and M.~Shigemori,
``On the matter of the Dijkgraaf-Vafa conjecture,''
JHEP {\bf 0304}, 052 (2003)
[arXiv:hep-th/0303104].

\bibitem{Argurio:2003ym}
R.~Argurio, G.~Ferretti and R.~Heise,
``An introduction to supersymmetric gauge theories and matrix models,''
[arXiv:hep-th/0311066].

\bibitem{seiberg02}
N.~Seiberg,
 ``Adding fundamental matter to 'Chiral rings and anomalies in supersymmetric
gauge theory',''
JHEP {\bf 0301}, 061 (2003)
[arXiv:hep-th/0212225].

\bibitem{Svrcek:2003az}
P.~Svrcek,
``Chiral rings, vacua and gaugino condensation of supersymmetric gauge
theories,''
[arXiv:hep-th/0308037].

\bibitem{csw1}
F.~Cachazo, N.~Seiberg and E.~Witten,
``Phases of N = 1 supersymmetric gauge theories and matrices,''
JHEP {\bf 0302}, 042 (2003)
[arXiv:hep-th/0301006].

\bibitem{CDSW}
F.~Cachazo, M.~R.~Douglas, N.~Seiberg and E.~Witten,
``Chiral rings and anomalies in supersymmetric gauge theory,''
JHEP {\bf 0212}, 071 (2002)
[arXiv:hep-th/0211170].

\bibitem{APS}
P.~C.~Argyres, M.~R.~Plesser and N.~Seiberg,
``The Moduli Space of N=2 SUSY {QCD} and Duality in N=1 SUSY {QCD},''
Nucl.\ Phys.\ B {\bf 471}, 159 (1996)
[arXiv:hep-th/9603042].

\bibitem{ckm}
G.~Carlino, K.~Konishi and H.~Murayama,
``Dynamical symmetry breaking in supersymmetric SU(n(c)) and USp(2n(c))  
gauge theories,''
Nucl.\ Phys.\ B {\bf 590}, 37 (2000)
[arXiv:hep-th/0005076].

\bibitem{ckkm}
G.~Carlino, K.~Konishi, S.~P.~Kumar and H.~Murayama,
``Vacuum structure and flavor symmetry 
breaking in supersymmetric  SO(n(c)) gauge theories,''
Nucl.\ Phys.\ B {\bf 608}, 51 (2001)
[arXiv:hep-th/0104064].

\bibitem{bfhn}
V.~Balasubramanian, B.~Feng, M.~x.~Huang and A.~Naqvi,
``Phases of N = 1 supersymmetric gauge theories with flavors,''
Annals Phys.\  {\bf 310}, 375 (2004)
[arXiv:hep-th/0303065].

\bibitem{afo1}
C.~Ahn, B.~Feng and Y.~Ookouchi,
``Phases of N = 1 SO(N(c)) gauge theories with flavors,''
Nucl.\ Phys.\ B {\bf 675}, 3 (2003)
[arXiv:hep-th/0306068].


\bibitem{afo2}
C.~Ahn, B.~Feng and Y.~Ookouchi,
``Phases of N = 1 USp(2N(c)) gauge theories with flavors,''
Phys.\ Rev.\ D {\bf 69}, 026006 (2004)
[arXiv:hep-th/0307190].

\bibitem{DJ}
Y.~Demasure and R.~A.~Janik,
 ``Explicit factorization of Seiberg-Witten curves with matter from random
matrix models,''
Nucl.\ Phys.\ B {\bf 661}, 153 (2003)
[arXiv:hep-th/0212212].

\bibitem{krs}
P.~Kraus, A.~V.~Ryzhov and M.~Shigemori,
``Loop equations, matrix models, and 
N = 1 supersymmetric gauge theories,''
JHEP {\bf 0305}, 059 (2003)
[arXiv:hep-th/0304138].





\bibitem{ac}
L.~F.~Alday and M.~Cirafici,
``Effective superpotentials via Konishi anomaly,''
JHEP {\bf 0305}, 041 (2003)
[arXiv:hep-th/0304119].



\bibitem{agaetal}
M.~Aganagic, K.~Intriligator, C.~Vafa and N.~P.~Warner,
``The glueball superpotential,''
[arXiv:hep-th/0304271].

\bibitem{cach}
F.~Cachazo,
``Notes on supersymmetric 
Sp(N) theories with an antisymmetric tensor,''
[arXiv:hep-th/0307063].

\bibitem{Matone:2003bx}
M.~Matone,
``The affine connection of supersymmetric SO(N)/Sp(N) theories,''
JHEP {\bf 0310}, 068 (2003)
[arXiv:hep-th/0307285].

\bibitem{ll}
K.~Landsteiner and C.~I.~Lazaroiu,
``On Sp(0) factors and orientifolds,''
Phys.\ Lett.\ B {\bf 588}, 210 (2004)
[arXiv:hep-th/0310111].

\bibitem{Naculich:2003ka}
S.~G.~Naculich, H.~J.~Schnitzer and N.~Wyllard,
``Matrix-model description of N = 2 gauge theories with non-hyperelliptic
Seiberg-Witten curves,''
Nucl.\ Phys.\ B {\bf 674}, 37 (2003)
[arXiv:hep-th/0305263].

\bibitem{Gomez-Reino:2004rd}
M.~Gomez-Reino, S.~G.~Naculich and H.~J.~Schnitzer,
``Improved matrix-model calculation of the N = 2 prepotential,''
JHEP {\bf 0404}, 033 (2004)
[arXiv:hep-th/0403129].

\bibitem{Strominger:1995cz}
A.~Strominger,
``Massless black holes and conifolds in string theory,''
Nucl.\ Phys.\ B {\bf 451}, 96 (1995)
[arXiv:hep-th/9504090].

\bibitem{IS-1}
K.~A.~Intriligator and N.~Seiberg,
``Phases of N=1 supersymmetric gauge theories in four-dimensions,''
Nucl.\ Phys.\ B {\bf 431}, 551 (1994)
[arXiv:hep-th/9408155].

\bibitem{Eli}
S.~Elitzur, A.~Forge, A.~Giveon, K.~A.~Intriligator and E.~Rabinovici,
``Massless Monopoles Via Confining Phase Superpotentials,''
Phys.\ Lett.\ B {\bf 379}, 121 (1996)
[arXiv:hep-th/9603051].

\bibitem{De}
Y.~Demasure,
``Affleck-Dine-Seiberg from Seiberg-Witten,''
[arXiv:hep-th/0307082].

\bibitem{EHU}
J.~Erlich, S.~Hong and M.~Unsal,
``Matrix models, monopoles and modified moduli,''
[arXiv:hep-th/0312054].

\bibitem{Seiberg:1994pq}
N.~Seiberg,
``Electric - magnetic duality in supersymmetric nonAbelian gauge theories,''
Nucl.\ Phys.\ B {\bf 435}, 129 (1995)
[arXiv:hep-th/9411149].

\bibitem{CIV}
F.~Cachazo, K.~A.~Intriligator and C.~Vafa,
``A large N duality via a geometric transition,''
Nucl.\ Phys.\ B {\bf 603}, 3 (2001)
[arXiv:hep-th/0103067].


\bibitem{Ookouchi}
Y.~Ookouchi,
``N = 1 gauge theory with flavor from fluxes,''
JHEP {\bf 0401}, 014 (2004)
[arXiv:hep-th/0211287].

\bibitem{Witten:1998xy}
E.~Witten,
``Baryons and branes in anti de Sitter space,''
JHEP {\bf 9807}, 006 (1998)
[arXiv:hep-th/9805112].


\bibitem{ao9}
C.~Ahn and Y.~Ookouchi,
``The matrix model curve near the singularities,''
[arXiv:hep-th/0309156].

\bibitem{NSW02}
S.~G.~Naculich, H.~J.~Schnitzer and N.~Wyllard,
``Matrix model approach to the N = 2 U(N) gauge theory with matter in the
fundamental representation,''
JHEP {\bf 0301}, 015 (2003)
[arXiv:hep-th/0211254].

\bibitem{CV}
F.~Cachazo and C.~Vafa,
``N = 1 and N = 2 geometry from fluxes,''
[arXiv:hep-th/0206017].















































































\end{thebibliography}
\end{document}